\documentclass[paper]{myJHEP3}
\newcommand{\feynSlash}[1]{\slashed{#1}}

\newcommand{\dd}{\mathrm{d}}
\newcommand{\FDRMeasure}[1]{[\dd{}^4 #1]}

\newcommand{\FDRbar}[1]{\overline{#1}}
\usepackage{axodraw4j}
\usepackage{amsmath}
\usepackage{slashed}
\usepackage[english]{babel}
\usepackage{cite}
\newcommand{\DRFDR}{{\rm D_{\rm {\scriptscriptstyle R}eg}}^{\rm\!\!\!\!\!\!\!\! \scriptscriptstyle FDR}}
\newcommand{\FDRFDH}{{\rm FDH}^\prime}

\newcommand{\beq}{\begin{equation}}
\newcommand{\eeq}{\end{equation}}
\newcommand{\bqa}{\begin{eqnarray}}
\newcommand{\eqa}{\end{eqnarray}}

\newcommand{\nl}{\nonumber \\}

\newcommand{\qbar}{\bar q}

\def\spa#1.#2{\left\langle#1\,#2\right\rangle}
\def\spb#1.#2{\left[#1\,#2\right]}

\def\feynsl#1{
  \setbox0=\hbox{/} \setbox1=\hbox{$#1$}
  \dimen0=\wd0 \advance\dimen0 by -\wd1 \divide\dimen0 by 2
  \ifdim\wd0>\wd1 \raise.15ex\copy0\kern-\wd0\kern\dimen0\copy1\kern\dimen0
  \else \kern-\dimen0\raise.15ex\copy0\kern-\dimen0\kern-\wd1\copy1\fi}

\newcommand\sss{\scriptscriptstyle}

\newcommand\mur{\mu_{\sss\rm R}}

\newskip\humongous \humongous=0pt plus 1000pt minus 100pt

\newif\ifdtup

\def    \br(#1,#2)          {\mbox{$\langle #1 \, #2 \rangle$}}
\def    \sq(#1,#2)          {\mbox{$\left[  #1 \, #2 \right]$}}

\title{Two-loop off-shell QCD amplitudes in FDR}
\author{Ben Page and Roberto Pittau\\
        CERN, PH-TH, Geneva, Switzerland {\rm and}\\ Departamento de F\'isica Te\'orica y del Cosmos and CAFPE,
  Campus Fuentenueva s.n., Universidad de Granada, E-18071 Granada, Spain \\
  E-mail: \email{ben.page@cern.ch,pittau@ugr.es}}

\abstract{We link the FDR treatment of ultraviolet (UV) divergences to dimensional regularization up to two loops in QCD. This allows us to derive
the one-loop and two-loop coupling constant and quark mass shifts necessary to translate infrared finite quantities computed in FDR to the $\overline{\rm MS}$
renormalization scheme.
As a by-product of our analysis, we solve a problem analogous to the breakdown of unitarity in the Four Dimensional Helicity (FDH) method beyond one loop. A fix
to FDH is then presented that preserves the renormalizability properties of QCD without introducing evanescent quantities.}
\preprint{}

\begin{document}
\section{Introduction}
When using customary approaches~\cite{Bogoliubov:1957gp,Hepp:1966eg,Zimmermann:1969jj,'tHooft:1972fi,Collins:1984xc} to handle the ultraviolet (UV) problem in quantum field theory (QFT), intermediate steps are necessary to extract physical answers from loop calculations.
In particular, the Lagrangian $\cal{L}$ of the theory is modified by adding UV counterterms (CTs). They absorb the divergences generated by the high-frequency part of the loop integrals and as one moves up the orders in the loop expansion one must include lower loop CT calculations in order to be consistent.
When carrying out this renormalization program UV infinities are usually regulated via dimensional regularization~\cite{'tHooft:1972fi} (DReg) and renormalized quantities are defined by specifying what is subtracted from the  bare ones. A particularly convenient subtraction scheme is $\overline{\rm MS}$, in which only UV poles and universal constants are dropped.

 The FDR\footnote{Acronym of Four Dimensional Regularization/Renormalization.} approach of~\cite{Pittau:2012zd} deals with UV infinities in a different way. A new kind of  loop integration\footnote{Called FDR integration.} is introduced that coincides with Riemann integration in UV finite cases, but produces a finite and regulator free answer also when acting on divergent integrands.\footnote{Abusing a bit the language, we dub divergent(convergent) {\em integrands} those which would generate UV divergent(convergent) {\em integrals} upon normal, four-dimensional integration.} In this way no CTs need to be incorporated into $\cal{L}$: they are traded for a change in the {\em definition} of the loop integration. Moreover, FDR directly generates renormalized amplitudes since it is  independent of any UV cutoff.

The main aim of this paper is to construct the one- and two-loop transition rules between FDR and $\overline{\rm MS}$ in the framework of QCD.
As a perturbative treatment of a renormalizable QFT is unique up to a renormalization scheme dependence,
we have achieved this task by studying 
{\em off-shell}\,\footnote{Working off-shell allows us to deal with IR convergent integrals. A detailed two-loop study of FDR in the presence of IR divergences is equally important, but outside our focus.} $\ell$-loop 
QCD correlation functions $G^{\ell}$ and by searching for a DReg renormalization scheme which reproduces the FDR answer when $\ell= 1$ and $\ell= 2$, i.e.
\begin{subequations}
\label{eq:fond}
\bqa
\label{eq:fond1}
G^{\ell}_{\rm renormalized} &=& G^{\ell}_{\rm bare} +(\ell\text{-loop-CTs})+
\ldots + (1\text{-loop-CTs}),\\ 
\label{eq:fond2}
G^{\ell}_{\rm renormalized} &=& G^{\ell}_{\rm FDR}.
\eqa
\end{subequations}
We dub such a scheme $\DRFDR$ and its renormalization constants 
${Z}^{\ell,{\rm \scriptscriptstyle FDR}}$. These can then be used to extract the coupling constant and quark mass shifts that relate FDR to $\overline{\rm MS}$.
We emphasize that in a typical FDR calculation {\em there are no} ${Z}^{\ell,{\rm \scriptscriptstyle FDR}}$, since one directly computes eq.~(\ref{eq:fond2}). We introduce them in the context of this paper only because we want to work out the correspondence of FDR with a canonical renormalization approach based on counterterms.

The second insight of our work concerns the nature of self contractions of Lorentz indices and $\gamma$-matrices explicitly appearing in $G^{\ell}_{\rm bare}$, that we denote by $n_s$. When studying eq.~(\ref{eq:fond}) one needs to
reconcile the value $n_s=n$ dictated by DReg in (\ref{eq:fond1}) with $n_s= 4$ used in (\ref{eq:fond2}). This has to be done without spoiling the renormalizability of QCD. Following FDR as a guidance, we have been able to establish the correct $n_s \to 4$ limit of DReg, compatible with a local subtraction parametrized in terms of the ${Z}^{\ell,{\rm \scriptscriptstyle FDR}}$. 
A mismatch between the structure of $G^{\ell}_{\rm bare}$ and the CTs has been recently recognized~\cite{Kilgore:2011ta,Boughezal:2011br,Kilgore:2012tb} to cause the failure of the renormalization program beyond one-loop in the original formulation of the Four Dimensional Helicity scheme (FDH)~\cite{Bern:1991aq,Bern:2002zk}. Since we observe that terms restoring the right cancellations are produced by FDR, we argue that the FDR approach provides a natural solution to this problem. In contrast with methods that require the introduction of $\epsilon$-scalars~\cite{Broggio:2015ata}, in our case no
new fields nor evanescent couplings need to be added to the QCD Lagrangian. Moreover, a shift linking the $\overline{\rm MS}$ quark mass to a fixed version of FDH can be worked out.  So far we have tested this approach for off-shell correlation functions. What happens in the presence of IR divergences needs to be further investigated.

The outline of the paper is as follows. In the next section we review FDR. Sections~\ref{sec:footing} and \ref{sec:calculation} describe our computational strategy.  Our results are collected in
section~\ref{sec:res} and appendix~\ref{sec:app0}. Furthermore, in section~\ref{sec:fdh} we discuss our fix to FDH.
Finally, appendices~\ref{sec:ex}--\ref{app:exx} contain explicit examples of our algorithm. 

\section{FDR} 
\label{FDR}
FDR integration was first introduced in~\cite{Pittau:2012zd} and several examples of one- and two-loop computations 
can be found in~\cite{Donati:2013iya,Donati:2013voa,Pittau:2013qla,Pittau:2013ica,Pittau:2014tva}. Here we briefly review its definition, which we need to make contact with our calculation.

Consider an $\ell$-loop integrand $J(q_1,\ldots,q_\ell)$ depending on $\ell$ loop
momenta $q_1,\ldots,q_\ell$. The multi-loop FDR integration over $J$ is defined
as\footnote{FDR integration is denoted by the symbol $[d^4q_i]$.}
\bqa
\label{eq:FDRdef}
\int [d^4 q_1] \cdots [d^4 q_\ell]\, J(q_1,\ldots,q_\ell,\mu^2) \equiv
\lim_{\mu \to 0} 
\int d^4 q_1 \cdots d^4 q_\ell \,
J_{\rm F}(q_1,\ldots,q_\ell,\mu^2),
\eqa
where $J_{\rm F}(q_1,\ldots,q_\ell,\mu^2)$ is the UV finite part of 
$J(q_1,\ldots,q_\ell,\mu^2)$, specified below, and $\mu$ 
a vanishing mass required to extract  $J_{\rm F}$ from $J$. 
$J(q_1,\ldots,q_\ell,\mu^2)$ and $J_{\rm F}(q_1,\ldots,q_\ell,\mu^2)$ are read off from $J(q_1,\ldots,q_\ell)$ according to the following rules:
\begin{itemize}
\item[i)] Squares of integration momenta appearing {\em both} in the denominators of
$J(q_1,\ldots,q_\ell)$ {\em and} in contractions generated by Feynman rules are shifted by $\mu^2$
\bqa
\label{eq:shift}
q^2_i \to q^2_i -\mu^2 \equiv \qbar^2_i.
\eqa
This replacement is called {\em global prescription};
\item[ii)] A splitting
\bqa
\label{eq:split}
J(q_1,\ldots,q_\ell,\mu^2)= [J_{\rm INF}(q_1,\ldots,q_\ell,\mu^2)]+
J_{\rm F}(q_1,\ldots,q_\ell,\mu^2)
\eqa
is performed in such a way that UV divergences are entirely parametrized in terms of divergent integrands\footnote{By convention, we write divergent integrands between square brackets and call them {\em FDR vacua}, or simply {\em vacua}.
Examples of the extraction of FDR vacua from loop integrands 
are given in appendix~\ref{sec:ex}.}  contained in $[J_{\rm INF}]$, that {\em solely depend on $\mu^2$};
\item[iii)]  The global prescription of eq.~(\ref{eq:shift}) should be made compatible with
a key property of multi-loop calculus:\footnote{This last requirement turns out to be the mechanism that enforces the renormalizability of the $\DRFDR$ scheme. It is discussed at length in section~\ref{sec:sub-bar}.}
\bqa
\label{eq:subcon}
\begin{tabular}{c}
{\em in an $\ell$-loop diagram,
one should be able to calculate a sub-diagram,} \\
{\em insert the integrated
form into the full diagram and get the same answer.} \\
\end{tabular}
\eqa
We dub this {\em sub-integration consistency}.
\end{itemize}
Finally, after $\lim_{\mu \to 0}$ is taken, $\ln \mu \to \ln \mur$ is understood in the r.h.s. of eq.~(\ref{eq:FDRdef}), where $\mur$ is an arbitrary renormalization scale. 

This definition preserves shift invariance
\bqa
\label{eq:shiftinv}
\int [d^4q_1] \ldots [d^4q_\ell]\,  J(q_1,\ldots, q_\ell,\mu^2) =\int [d^4q_1] \ldots [d^4q_\ell]\,  J(q_1+p_1,\ldots, q_\ell+p_\ell,\mu^2), 
\eqa
and the possibility of canceling numerators and denominators
\bqa
\label{eq:canc}
\int [d^4q_1] \ldots [d^4q_\ell]\, \frac{\qbar^2_i-m^2_i}{ (\qbar^2_i-m^2_i)^m \ldots} = 
\int [d^4q_1] \ldots [d^4q_\ell]\, \frac{1}{ (\qbar^2_i-m^2_i)^{m-1}\ldots},
\eqa
which are properties needed to retain the symmetries of ${\cal L}$.
From eqs.~(\ref{eq:shiftinv}) and~(\ref{eq:canc}) it follows that algebraic manipulations in FDR integrands are allowed as if they where convergent ones. This authorizes one to reduce complicated multi-loop integrals to a limited set of Master Integrals (MI) by using four-dimensional tensor decomposition~\cite{Donati:2013voa} or integration-by-parts identities~\cite{Pittau:2014tva}. In other words, the definition in eq.~(\ref{eq:FDRdef}) can be applied just at the end of the calculation, when the actual value of the MIs is needed. 

An important subtlety implied by eq.~(\ref{eq:canc}) is that the needed cancellation works only if integrands involving explicit powers of $\mu^2$ in the numerator are also split via eq.~(\ref{eq:split}) {\em as if $\mu^2 = q^2_i$}, where $q^2_i$ is the momentum squared which generates $\mu^2$. As a consequence, although only one kind of $\mu^2$ exists, one has to keep track of its origin when it appears in the numerator of $J(q_1,\ldots,q_\ell,\mu^2)$. For this we use the notation  $\mu^2|_i$, which understands the same splitting required for $q^2_i$. 
FDR integrals with powers of $\mu^2|_i$ in the numerator are called ``extra integrals''. Their computation is elementary. One- and two-loop examples can be found in~\cite{Pittau:2012zd,Donati:2013voa} and appendix~\ref{app:exx}.

\section{FDR and DReg on the same footing} 
\label{sec:footing}
Eq.~(\ref{eq:fond}) defines the $\DRFDR$ scheme we are looking for, i.e. the
DReg scheme reproducing the FDR correlation functions. As the r.h.s. of~(\ref{eq:fond1}) is computed in DReg while the r.h.s.
of~(\ref{eq:fond2}) in FDR, we need a common framework  which accommodates both approaches.
Here we illustrate this framework at two loops, but the same considerations apply at any loop order.

Our starting point is eq.~(\ref{eq:split}) when $J(q_1,q_2)$ is the sum of all integrands contributing to $G^2_{\rm bare}$.
{\em If} $J(q_1,q_2)$ is free of IR infinities we can rewrite the DReg integration over it as\footnote{$n$ is the number of dimensions defined as $n= 4-2\epsilon$.}
\bqa
\label{eq:strat}
&&\int d^n q_1 d^n q_2\, J(q_1,q_2) 
= 
\int d^n q_1 d^n q_2\, \lim_{\mu \to 0} J(q_1,q_2,\mu^2)=
\lim_{\mu \to 0} \int d^n q_1 d^n q_2\, J(q_1,q_2,\mu^2) \nl
&&~=
\lim_{\mu \to 0} 
\int d^n q_1 d^n q_2\, \left[J_{\rm INF}(q_1,q_2,\mu^2)\right]+
\lim_{\mu \to 0} 
\int d^4 q_1 d^4 q_2\, J_{\rm F}(q_1,q_2,\mu^2),
\eqa
where the absence of IR divergences authorizes us to extract $\lim_{\mu \to 0}$ out of the integral. 
The integral over $J_{\rm F}(q_1,q_2,\mu^2)$ is $G^2_{\rm FDR}$ and through
eq.~(\ref{eq:strat}) we have isolated it within $G^2_{\rm bare}$, so that the $\DRFDR$ scheme can be determined by solely looking at the pieces which differ between eq.~(\ref{eq:fond1}) and~(\ref{eq:fond2}), namely
\bqa
\left[J_{\rm INF}(q_1,q_2,\mu^2)\right],~~(1\text{-loop-CTs})~~{\rm and}~~(2\text{-loop-CTs}).
\eqa
This defines our strategy. For instance  the $(1\text{-loop-CTs})$ are known at two loops, so that $\DRFDR$ is defined by choosing
\bqa
\label{eq:2ct}
 (2\text{-loop-CTs})= -(1\text{-loop-CTs})
-\lim_{\mu \to 0} \int d^n q_1 d^n q_2\, \left[J_{\rm INF}(q_1,q_2,\mu^2)\right],
\eqa
which sets the finite part of $G^2_{\rm bare}$ to the FDR value. 

An important condition should be fulfilled by eq.~(\ref{eq:2ct}). It should be {\em local} and proportional to the the Born correlation 
function $G^0$:
\bqa
\label{eq:rencon}
(2\text{-loop-CTs})= Const  \times G^0,
\eqa
which is the correct form to be re-absorbed into ${Z}^{2,{\rm \scriptscriptstyle FDR}}$.
We dub eq.~(\ref{eq:rencon}) our {\em renormalizability condition}.
To understand its consequences it is convenient to split $\left[J_{\rm INF}\right]$ as follows:
\bqa
\label{eq:splitdiv}
\left[J_{\rm INF}(q_1,q_2,\mu^2)\right]= \left[J^{\scriptscriptstyle {\rm SV}}_{\rm INF}(q_1,q_2,\mu^2)\right]+\left[J^{\scriptscriptstyle {\rm GV}}_{\rm INF}(q_1,q_2,\mu^2)\right].
\eqa
$\left[J^{\scriptscriptstyle {\rm SV}}_{\rm INF}\right]$ is a factorizable contribution in which only one sub-integration is UV divergent. It is called the {\em sub-vacuum} and depends on the kinematic scales entering the UV finite sub-integration.
The second piece is independent of kinematics and it is called the {\em 
global vacuum.}\footnote{Examples of global vacua and sub-vacua are given in the first line of eq.~(\ref{eq:exglo}) and in eq.~(\ref{eq:j7}), respectively.}
Eqs.~(\ref{eq:2ct}) and (\ref{eq:rencon}) require that, upon integration, kinematic dependent terms in $\left[J^{\scriptscriptstyle {\rm SV}}_{\rm INF}\right]$ should cancel one-loop counterterms. In section~\ref{sec:calculation} we  will describe the subtleties of this cancellation. 
\section{The calculation} 
\label{sec:calculation}
As outlined in the previous section the core of our strategy is calculating FDR vacua in DReg instead of throwing them away. As a first step we write down\footnote{We work in the Feynman--'t Hooft gauge using {\tt QGRAF}~\cite{Nogueira:1991ex} for generating the diagrams and {\tt FORM}~\cite{Kuipers:2012rf} for extracting their vacuum part.} the twelve integrands $J^{G^{\ell}_i}$ ($i= 1,\ldots,6$ and $\ell = 1,2$) corresponding to the one-particle irreducible one- and two-loop QCD correlation functions drawn in figure~\ref{fig:1}, excluding CT diagrams. Then, we split them as in eq.~(\ref{eq:split}) and compute the integrals over their vacuum parts, 
\bqa
\label{eq:ourJ}
I^{1}_{i}= \lim_{\mu \to 0} 
\int \frac{d^n q}{\mur^{-2 \epsilon}}
\, \left[J^{G^{1}_i}_{\rm INF}(q,\mu^2)\right]~~~{\rm and}~~~
I^{2}_{i}= \lim_{\mu \to 0} 
\int \frac{d^n q_1}{\mur^{-2 \epsilon}} \frac{d^n q_2}{\mur^{-2 \epsilon}}
\, \left[J^{G^{2}_i}_{\rm INF}(q_1,q_2,\mu^2)\right],
\eqa
in the symmetric off-shell kinematic point
\bqa
\label{eq:offkin}
k^2_1 =k^2_2 =k^2_3 = M^2.
\eqa
With this choice all integrals are made free of IR divergences, so eq.~(\ref{eq:strat}) applies. Furthermore, the complexity of the calculation is reduced to a one-scale problem.
\FIGURE{
\begin{picture}(300,190)(0,0)
\SetScale{1}
\SetWidth{0.5}
\SetOffset(-90,160)

\Gluon(45,0)(90,0){4}{4}
\Gluon(90,0)(135,0){4}{4}
\GCirc(90,0){12}{0.8}
\Text(65,15)[b]{$k_1$}
\LongArrow(55,9)(70,9)
\Text(90,-25)[t]{\large $G^{\ell}_1= G^{\ell}_{GG}$}
\SetOffset(60,160)
\DashArrowLine(45,0)(90,0){2.5}
\DashArrowLine(90,0)(135,0){2.5}
\GCirc(90,0){12}{0.8}
\Text(65,15)[b]{$k_1$}
\LongArrow(55,9)(70,9)
\Text(90,-25)[t]{\large $G^{\ell}_2= G^{\ell}_{cc}$}
\SetOffset(210,160)
\ArrowLine(45,0)(90,0)
\ArrowLine(90,0)(135,0)
\GCirc(90,0){12}{0.8}
\Text(65,15)[b]{$k_1$}
\LongArrow(55,9)(70,9)
\Text(90,-25)[t]{\large $G^{\ell}_3= G^{\ell}_{\Psi\Psi}$}
\SetOffset(-90,70)
\Gluon(45,25)(90,0){4}{4}
\Gluon(90,0)(135,25){4}{4}
\Gluon(90,0)(90,-40){4}{4}
\GCirc(90,0){12}{0.8}
\Text(67,30)[b]{$k_1$}
\LongArrow(55,30)(70,22)
\Text(110,30)[b]{$k_2$}
\LongArrow(124,30)(109,22)
\Text(104,-28)[l]{$k_3$}
\LongArrow(99,-34)(99,-20)
\Text(90,-53)[t]{\large $G^{\ell}_4= G^{\ell}_{GGG}$}
\SetOffset(60,70)
\DashArrowLine(45,25)(90,0){2.5}
\DashArrowLine(90,0)(135,25){2.5}
\Gluon(90,0)(90,-40){4}{4}
\GCirc(90,0){12}{0.8}%
\Text(67,30)[b]{$k_1$}
\LongArrow(55,30)(70,22)
\Text(110,30)[b]{$k_2$}
\LongArrow(124,30)(109,22)
\Text(104,-28)[l]{$k_3$}
\LongArrow(99,-34)(99,-20)
\Text(90,-53)[t]{\large $G^{\ell}_5= G^{\ell}_{Gcc}$}

\SetOffset(210,70)
\ArrowLine(45,25)(90,0)
\ArrowLine(90,0)(135,25)
\Gluon(90,0)(90,-40){4}{4}
\GCirc(90,0){12}{0.8}
\Text(67,30)[b]{$k_1$}
\LongArrow(55,30)(70,22)
\Text(110,30)[b]{$k_2$} 
\LongArrow(124,30)(109,22)
\Text(104,-28)[l]{$k_3$}
\LongArrow(99,-34)(99,-20)
\Text(90,-53)[t]{\large $G^{\ell}_6= G^{\ell}_{G\Psi\Psi}$}
\end{picture}
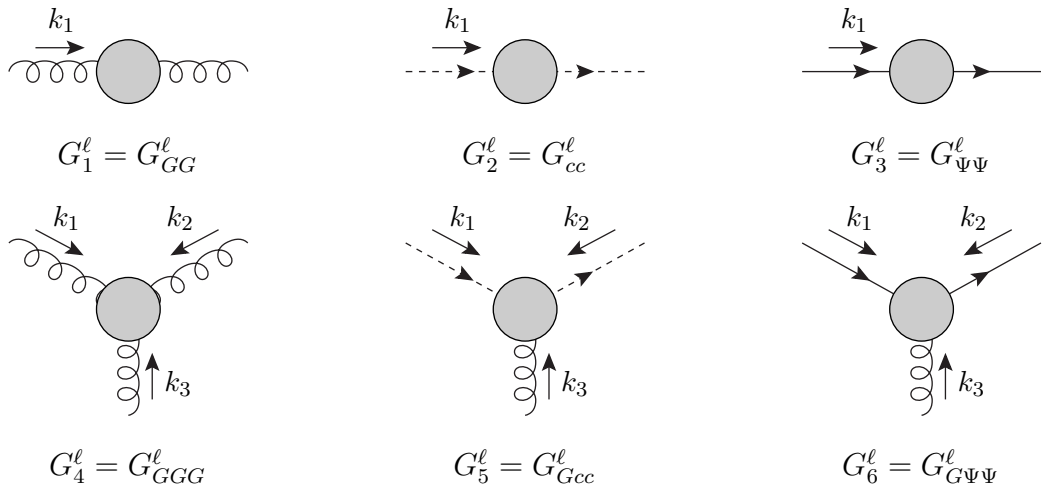
\caption{One-particle irreducible one-loop ($\ell = 1$) and two-loop ($\ell = 2$) QCD Green's functions used 
in our calculation. The gray blobs denote the sum of all possible Feynman diagrams computed in the Feynman--'t Hooft gauge. Diagrams with counterterms {\em are not} included when the $G^{\ell}_i$ are used to compute the r.h.s. of eq.~(\ref{eq:fond2}).} 
\label{fig:1}
}
An important point concerns explicit contractions appearing in the numerator of $\left[J^{G^{\ell}_i}_{\rm INF}(q,\mu^2)\right]$:
\bqa
\label{eq:ns}
\gamma^\alpha  \gamma_\alpha = g^{\alpha\beta}g_{\alpha\beta}= n_s.
\eqa
As $I^{1}_{i}$ and $I^{2}_{i}$ are regulated in DReg, $n_s=n$ should be used in
eq.~(\ref{eq:ourJ}).

The knowledge of the $I^{\ell}_{i}$ allows us to parametrize the FDR subtraction in terms of renormalization constants $Z^{\ell,\rm FDR}_i$. However, not all of them are independent, as they are related by QCD Slavnov-Taylor (S-T) identities.\footnote{See section \ref{sec:res}.}
In the following we consider, in turn, the
$\ell = 1$ and $\ell = 2$ contributions to the $Z^{\ell,\rm FDR}_i$.

\subsection{The one-loop case}
At one loop the calculation of $I^{1}_{i}$ is simple. Schematically:
\begin{itemize}
\item The global prescription of eq.~(\ref{eq:shift}) ensures that reducible  $\qbar^2$s in the numerator simplify with denominators; 
\item The integrands are then split as in eq.~(\ref{eq:split}) by using the propagator identity
\bqa
\frac{1}{\qbar^2+2(q \cdot k) + k^2}= 
\frac{1}{\qbar^2} - 
\frac{2(q \cdot k) + k^2}{\qbar^2(\qbar^2+2(q \cdot k) + k^2)},
\eqa
which allows one to express $I^{1}_{i}$ in terms of tensors of the kind
\bqa
\int \frac{d^n q}{\mur^{-2 \epsilon}} \frac{q^{\alpha_1} \cdots q^{\alpha_{2r}}}{(q^2-\mu^2)^{s}},~~~~~(4+2r-2s \ge 0);
\eqa
\item
Finally, by virtue of $\lim_{\mu \to 0}$,  polynomially divergent integrals vanish, so any tensor structure is reduced to a fundamental scalar
\bqa
\int \frac{d^n q}{\mur^{-2 \epsilon}} \frac{q^{\alpha_1} \cdots q^{\alpha_{2r}}}{(q^2-\mu^2)^{r+2}}
= \frac{2}{(2r+2)!!} g^{\alpha_1 \cdots \alpha_{2r}} V(\mu),
\eqa
with
\bqa
\label{eq:simple}
V(\mu) &\equiv& \int \frac{d^n q}{\mur^{-2 \epsilon}} \frac{1}{(q^2-\mu^2)^2}
= i \pi^2 (\Delta-\ln \frac{\mu^2}{\mur^2}) + {\cal O}(\epsilon), \\
\Delta &=& \frac{1}{\epsilon}-\gamma_E -\ln \pi,
\eqa
and where $g^{\alpha_1 \cdots \alpha_{2r}}$ is completely symmetric and only made of products of metric tensors.
\end{itemize}
Eventually, the logarithm in eq.~(\ref{eq:simple}) has to be combined with the one-loop analogue of the $J_{\rm F}$ term in eq.~(\ref{eq:strat}) to compensate the $\mu$ dependence of the finite part.\footnote{This is why FDR integration is defined with the replacement $\ln \mu \to \ln \mur$.} As a result, the UV divergent part of $I^1_i$, ${\rm Inf}(I^1_i)$, is fully proportional to\footnote{The notation $\left.\right|_{\epsilon \to 0}$ means neglecting terms of ${\cal O}(\epsilon)$.}
\bqa
\label{eq:V0}
V_0= \left.V(\mur)\right|_{\epsilon \to 0}= i \pi^2\Delta.
\eqa
Now we study ${\rm Fin}(I^1_i)$, namely the UV convergent contribution to $I^1_i$. It determines the finite part of the one-loop subtraction term in eq.~(\ref{eq:fond1}) and  appears due to the explicit presence of $n_s$ in $\left[J^{G^{1}_i}_{\rm INF}(q,\mu^2)\right]$.
In fact, rewriting
\bqa
n_s= 4-2\lambda \epsilon,
\eqa
we see that when $\lambda \epsilon$ multiplies UV single poles, constants are generated that are fully subtracted in $\DRFDR$ but contribute to the finite part of $G^{1}_i$ in $\overline{\rm MS}$. Thus, the requirement of eq.~(\ref{eq:fond2}) causes a deviation from the minimality of the $Z^{1,\rm FDR}_i$ proportional to $(1-\lambda)$, such that $\lambda= 0$ (1) in $\DRFDR$ ($\overline{\rm MS}$). As one finds that both ${\rm Inf}(I^1_i)$ and ${\rm Fin}(I^1_i)$ factorize the Born, one computes
\bqa
Z^{1,\rm FDR}_i G^{0}_i= -I^1_i.
\eqa

In summary, at one loop it is possible to perform a calculation in DReg and consistently renormalize the result to reproduce the FDR answer. In other words, FDR can be interpreted as a particular renormalization scheme of DReg, i.e. the $\DRFDR$ scheme we are looking for.

\subsection{The two-loop case}
\label{sec-tloop}
Here we illustrate the calculation for massless QCD. When dealing with a non vanishing quark mass $m_q$ the formulae complicate a bit, but the reasoning remains unchanged. For simplicity, we start from cases\footnote{Specified later on.} in which we see that the sub-integration consistency~(\ref{eq:subcon}) does not play any role, and postpone to section~\ref{sec:sub-bar} the study of more complicated situations.

After scalarization by means of tensor reduction and integration-by-parts, one finds two types of contributions to $I^{2}_{i}$:
\begin{itemize}
\item An integral over the global vacuum of the kind
\bqa
\label{eq:globv}
GV_i(\mu)= \lim_{\mu\to 0}\int \frac{d^n q_1}{\mur^{-2 \epsilon}} \frac{d^n q_2}{\mur^{-2 \epsilon}} \left(F_{i1}(n,n_s) \left[\frac{1}{\qbar^4_1 \qbar^4_2}\right]+F_{i2}(n,n_s)
                          \left[\frac{1}{\qbar^4_1 \qbar^2_2 \qbar^2_{12}}\right] 
\right),
\eqa
where $F_{ij}(n,n_s)$ are rational functions and $q_{12}= q_1+q_2$;
\item An integral over the factorizable sub-vacuum\footnote{This sub-vacuum is generated by the application of the identities
(with $i \ne j$)
\bqa
\frac{1}{\qbar_i^2}  = \frac{1}{\qbar_j^2}  
                     \left( 1-\frac{q^2_{ij}
                          -2(q_j\cdot q_{ij})}{\qbar_i^2}
                                             \right),~~   
\frac{1}{\qbar_i^2}  = \frac{1}{\qbar_{ij}^2}\left(1+\frac{q^2_{j}
                        +2(q_i\cdot q_j)}{\qbar_i^2}
                                             \right),~~  
\frac{1}{\qbar_{ij}^2}= \frac{1}{\qbar_i^2} \left(1-\frac{q^2_{j} 
                        +2(q_i\cdot q_j)}{\qbar_{ij}^2}
                                          \right) \nl 
\eqa
needed to disentangle the sub-divergences of $I^{2}_{i}$ from the finite part.} 
 of the form
\bqa
\label{eq:subv}
SV_i(\mu)= \lim_{\mu\to 0}\int \frac{d^n q_1}{\mur^{-2 \epsilon}} \frac{d^n q_2}{\mur^{-2 \epsilon}}
\left[\frac{1}{\qbar^4_2} \right] J_i(q_1, \mu),
\eqa
in which $J_i(q_1,\mu)$ is UV convergent.
\end{itemize}
By judiciously using the identity
\bqa
\label{eq:invprop}
 \frac{1}{\qbar_1^2+2(q_1 \cdot k) + k^2}= \frac{1}{k^2}
-\frac{\qbar_1^2+2(q_1 \cdot k)}{k^2(\qbar_1^2+2(q_1 \cdot k) + k^2)}
\eqa
$J_i(q_1,\mu)$ 
can be split into a piece which develops a $\ln \mu^2$ upon integration and a term where $\mu$ can be set to zero at the integrand level
\bqa
\label{eq:jsplit1}
J_i(q_1,\mu)= J_{ai}(q_1,\mu)+J_{bi}(q_1).
\eqa
With our special kinematics we find
\bqa
\label{eq:jsplit2}
J_{ai}(q_1,\mu)&=& \frac{A_i(n,n_s)}{\qbar^4_1}, \nl
J_{bi}(q_1)      &=& \frac{A_i^\prime(n,n_s)}{q^2_1(q_1+k_1)^2}
+ M^2 \frac{B_i(n,n_s)}{q^2_1(q_1-k_1)^2(q_1+k_2)^2}, 
\eqa
where $A_i(n,n_s)$, $A_i^\prime(n,n_s)$ and $B_i(n,n_s)$ are rational functions.
Notice that the UV finiteness of $J_i(q_1,\mu)$ ensures that the  
the pole parts of the integrals over $J_{ai}(q_1,\mu)$ and $J_{bi}(q_1)$ cancel each other.
Furthermore, $B_i(n,n_s)= 0$ in the case of two-point correlation functions.
An explicit example of the procedure yielding eqs.~(\ref{eq:globv})--(\ref{eq:jsplit2}) is reported in appendix~\ref{sec:ex}.

The additional complexity at two loops is that, unlike
\bqa
\label{eq:subpiecea}
SV_{ai}(\mu)= \lim_{\mu\to 0}\int \frac{d^n q_1}{\mur^{-2 \epsilon}} \frac{d^n q_2}{\mur^{-2 \epsilon}}
\left[\frac{1}{\qbar^4_2} \right] J_{ai}(q_1,\mu),
\eqa
the integral
\bqa
\label{eq:subpieceb}
SV_{bi}= \lim_{\mu\to 0}\int \frac{d^n q_1}{\mur^{-2 \epsilon}} \frac{d^n q_2}{\mur^{-2 \epsilon}}
\left[\frac{1}{\qbar^4_2} \right] J_{bi}(q_1)
\eqa
depends on the kinematics. Thus, it generates logarithms of physical scales that cannot be absorbed into the $Z^{2,\rm FDR}_i$. Therefore eq.~(\ref{eq:2ct}) tells us that, in order to establish the connection between
FDR and the standard renormalization approach, we must demonstrate that such non-local terms are compensated by the sum of all diagrams containing insertions of one-loop counterterms.\footnote{For consistency with the form of the 
$Z^{1,\rm FDR}_i$ the one-loop insertions have to be computed with $\lambda= 0$.} 
This contribution is dubbed $CT_i$ in the following. To achieve this cancellation, we have to recast eq.~(\ref{eq:subpieceb}) 
into a form suitable to be combined with the $CT_i$. Observing that\footnote{$V_0$ is defined in eq.~(\ref{eq:V0}).}
\bqa
SV_i(\mu)= V_0 \lim_{\mu \to 0} \int \frac{d^n q_1}{\mur^{-2 \epsilon}} J_i(q_1, \mu)+ \cal{O}(\epsilon)
\eqa
leads us to consider
\bqa
\label{eq:subpiecep}
SV^\prime_{ai}(\mu)&=& V_0\lim_{\mu\to 0}\int \frac{d^n q_1}{\mur^{-2 \epsilon}} J_{ai}(q_1,\mu), \nl
SV^\prime_{bi}&=& V_0\lim_{\mu\to 0}\int \frac{d^n q_1}{\mur^{-2 \epsilon}} J_{bi}(q_1)
\eqa 
instead of eqs.~(\ref{eq:subpiecea}) and~(\ref{eq:subpieceb}). 
Now $SV^\prime_{bi}$ has the same structure of a counterterm diagram, as it
factorizes $V_0$. An explicit calculation of the $CT_i$ gives
\bqa
\label{eq:compen}
CT_i+SV_{bi}^\prime= 0~~{\rm for}~~i= 1,2,4,5,
\eqa
and we obtain, for correlation functions without external fermions, the 
following two-loop renormalization constants
\bqa
Z^{2,\rm FDR}_i G^0_i= -GV_i(\mur)-SV^\prime_{ai}(\mur),
\eqa
in accordance with the renormalizability condition of eq.~(\ref{eq:rencon}).
As in eq.~(\ref{eq:V0}), the choice of the point $\mu= \mur$ has the effect
of removing the dependence on $\mu$ from the last integral in eq.~(\ref{eq:strat}). 

As in the one-loop case, $n_s$ in eqs.~(\ref{eq:globv}) and~(\ref{eq:jsplit2}) generates finite terms proportional to $(1-\lambda)$ in the $Z^{2,\rm FDR}_i$. 
In addition, finite contributions are created by two-loop integrands that do not multiply powers of $\lambda \epsilon$, e.g.
\bqa
\left[\frac{f(n)}{\qbar^4_1 \qbar^2_2 \qbar^2_{12}}\right].
\eqa
To denote their origin we multiply by a parameter $\delta$ any finite combination
\bqa
                 \int d^n q_1 d^n q_2\,\left[\frac{f(n)}{\qbar^4_1 \qbar^2_2 \qbar^2_{12}}\right]-
{\rm Pole\,Part}\left\{\int d^n q_1 d^n q_2\,\left[\frac{f(n)}{\qbar^4_1 \qbar^2_2 \qbar^2_{12}}\right]\right\}, 
\eqa
which is fully subtracted in $\DRFDR$ but contributes to the finite part of
$G^2_{i}$ in $\overline{\rm MS}$, so that $\delta= 1$ (0) in $\DRFDR$ ($\overline{\rm MS}$).

\subsection{``Extra''-extra integrals and sub-prescription}
\label{sec:sub-bar}
In the case of QCD Green's functions with external fermions
eq.~(\ref{eq:compen}) does not hold true\footnote{Interestingly, it applies when computing $I^2_3$ and $I^2_6$ with $n_s= 4$, which is the value prescribed by FDH, as discusses in section~\ref{sec:fdh}. Nevertheless, we emphasize that we {\em must} set $n_s$ to $n$ because our strategy is to reproduce the FDR result by regularizing FDR vacua in DReg, which dictates $n_s= n$.} and leads to results incompatible with eq.~(\ref{eq:rencon}).
In this section, we show that the FDR sub-integration consistency~(\ref{eq:subcon}) requires the introduction in the finite part of the correlation 
functions of ``extra''-extra integrals ($EEI$s) with the same structure of $SV^\prime_i$, namely
\bqa
\label{eq:spliteei}
EEI_i(\mu)= EEI_{ai}(\mu)+EEI_{bi}. 
\eqa
They are just what is needed to restore 
the renormalizability condition, i.e.\footnote{When calculating the vacuum $EEI$s contribute with a minus sign and should be computed with $n_s= n$.} 
\bqa
\label{eq:fixcanc}
CT_i+SV_{bi}^\prime-EEI_{ai}(\mu)-EEI_{bi} = -EEI_{ai}(\mu) 
\eqa
and
\bqa
\label{eq:z2fdr}
Z^{2,\rm FDR}_i G^0_i = -GV_i(\mur)-SV^\prime_{ai}(\mur)+EEI_{ai}(\mur)~~~(i=3,6),
\eqa
with $EEI_{ai}$ independent of physical scales.

The FDR origin of the $EEI_i$ is the need of introducing a 
{\em sub-prescription} to cure the mismatch between the {\em global prescription} of eq.~(\ref{eq:shift}) and the consistency condition~(\ref{eq:subcon}).
In fact, although essential to preserve gauge invariance at two loops, the shift
\bqa
\label{eq:shift1}
q^2_i \to q^2_i-\mu^2|_i
\eqa 
may clash with the analogous replacement required to ensure~(\ref{eq:subcon}) at the level of divergent one-loop sub-diagrams. This is better explained with an example. Consider the two-loop diagram\footnote{This is the only possible contribution to $G^2_3$ proportional to $N_f$.} of figure~\ref{fig:loop1}.
\FIGURE{
\begin{picture}(300,77) (0,-20)
\SetOffset(0,-10)
\SetScale{0.5}
    \SetWidth{1.3}
    \SetColor{Black}
    \GluonArc(100,0)(80,0,180) {6} {18}
    
    \Line[arrow,arrowpos=0.5,arrowlength=7.5,arrowwidth=4,arrowinset=0.2](-20,0)(220,0)

    \COval(100,80)(26,26)(0){White}{White}
    \Arc[arrow,arrowpos=0.5,arrowlength=7.5,arrowwidth=4,arrowinset=0.2,flip](100,80)(26,0,-180)
    \Arc[arrow,arrowpos=0.5,arrowlength=7.5,arrowwidth=4,arrowinset=0.2,clock](100,80)(26,0,180)

    \Text(36.5,2)[lb]{$q_1\!+\!k_1$}
    \Text(96.5,5)[lb]{$\alpha$}
    \Text(-5,3)[lb]{$\beta$}
    \Text(7.5,37.5)[lb]{$q_1$}
    \Line[arrow,arrowpos=1.0,arrowlength=4,arrowwidth=4,arrowinset=0.2](42,78)(22,58)
    \Line[arrow,arrowpos=1.0,arrowlength=4,arrowwidth=4,arrowinset=0.2](178,58)(158,78)
    \Text(85,37.5)[lb]{$ q_1$}
    \Text(45,57.5)[lb]{$q_2$}
    \Text(45,17.5)[lb]{$q_{12}$}
    \Text(120,22)[l]{${\displaystyle =~\int [d^4q_1] [d^4q_2] 
\frac{N(q_1,q_2)}{\qbar^4_1\qbar^2_2 \qbar^2_{12} (\qbar^2_1+k_1^2+2 (q_1 \cdot k_1))}}$}
\end{picture}
\caption{The two-loop diagram contributing the $N_f$ corrections to $G^2_3$.
$N(q_1,q_2)$ is given in the text dropping irrelevant constants.
The replacement $q^2_i \to \qbar^2_i$ is performed only in the denominators.
}
\label{fig:loop1}
}
Before applying eq.~(\ref{eq:shift1}) its numerator reads
\bqa
N(q_1,q_2)= 4 \gamma_\alpha (\rlap/q_1 + \rlap/k_1) \gamma_\beta
\left(
-g^{\alpha \beta} (q_2 \cdot q_{12}) + q_{12}^\alpha q_2^\beta + q_{12}^\beta q_2^\alpha
\right).
\eqa
The sub-prescription is defined as the effect of eq.~(\ref{eq:shift1}) on $N(q_1,q_2)$ from the point of view of the divergent sub-diagram pulled out from the rest of the diagram, as in figure~\ref{fig:loop1cut}. 
\FIGURE{
\begin{picture}(200,75) (0,-20)
\SetOffset(40,-15)
\SetScale{0.5}
    \SetWidth{1.3}
    \SetColor{Black}
    \GluonArc(100,0)(80,0,180) {6} {18}
    
    \Line[arrow,arrowpos=0.5,arrowlength=7.5,arrowwidth=4,arrowinset=0.2](-20,0)(220,0)

    \COval(100,80)(26,26)(0){White}{White}
    \Arc[arrow,arrowpos=0.5,arrowlength=7.5,arrowwidth=4,arrowinset=0.2,flip](100,80)(26,0,-180)
    \Arc[arrow,arrowpos=0.5,arrowlength=7.5,arrowwidth=4,arrowinset=0.2,clock](100,80)(26,0,180)

    \CBox(15,32)(43,55){White}{White}
    \SetWidth{2.0}
        
    \put(5,16)
    {
        \Line(0,26)(26,0)
        \Line(6,31)(32,5)
    }
    \CBox(185,32)(157,55){White}{White}
        
    \put(82,16)
    {
        \Line(26,26)(0,0)
        \Line(20,31)(-6,5)
    }

    \Text(23.5,39)[lb]{$ \beta$}
    \Text(72.5,42)[lb]{$ \alpha$}
    \Text(96.5,5)[lb]{$\hat \alpha$}
    \Text(-5,3)[lb]{$\hat \beta$}

\end{picture}
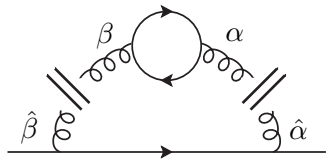
\caption{The same diagram of figure~\ref{fig:loop1} from the point of view of the $q_2$ sub-integration. Lorentz indices external to the the sub-diagram are given a hat.}
\label{fig:loop1cut}
}
From the perspective of the $q_2$ sub-integration one has to distinguish internal and external parts.
We denote the separation of the parts “external” to the sub-diagram by
placing hats on the Lorentz indices. Algebraically the hats do not make any difference other than to denote the fact that they are of an origin which is external to the sub-diagram, so all standard identities apply, for example
\bqa
{ \gamma}_{\hat \alpha}
{ \gamma}^{\hat \alpha}
=
{\gamma}_{\alpha}
{\gamma}^{\hat \alpha}
= 4,~~~
{\gamma}_{\hat \alpha}
{q_2}^\alpha= \rlap / {\hat q_2}. 
\eqa
We then study how the part\footnote{Here this is quadratic because less-than-quadratic terms do not generate extra integrals.} of $N(q_1,q_2)$ divergent in $q_2$
\bqa
N^{(2)}(q_1,q_2)=
8(\rlap/q_1 + \rlap/k_1)q^2_2+
8\rlap /{\hat q}_2(\rlap/q_1 + \rlap/k_1)\rlap /{\hat q}_2,
\eqa 
transforms under eq.~(\ref{eq:shift1}).
From the perspective of the sub-prescription the second term is inert, since it has an origin external to the sub-diagram. Thus
\bqa
\label{eq:localprescr}
N^{(2)}(q_1,q_2) \to N^{(2)}(q_1,q_2)
-8(\rlap/q_1 + \rlap/k_1)\mu^2|_2.
\eqa
On the other hand, the two-loop global prescription dictates 
that also the second term must transform as\footnote{This is obtained by anticommuting until the $\hat q_2$ meet.}
\bqa
8\rlap /\hat q_2(\rlap/q_1 + \rlap/k_1)\rlap /\hat q_2 \to
8\rlap /\hat q_2(\rlap/q_1 + \rlap/k_1)\rlap /\hat q_2
+8(\rlap/q_1 + \rlap/k_1)\mu^2|_2,
 \eqa
leading to no change in $N^{(2)}$:
\bqa
\label{eq:globalprescr}
N^{(2)}(q_1,q_2) \to N^{(2)}(q_1,q_2).
\eqa
Therefore, to remove double counting, the $EEI$ to be added to the diagram in figure~\ref{fig:loop1} is defined as the result of the sub-prescription minus
the outcome of the global prescription, i.e. the difference\footnote{By doing so, we reintroduce the correct one-loop behaviour leading to~(\ref{eq:subcon}).} between the r.h.s. of eqs.~(\ref{eq:localprescr}) and~(\ref{eq:globalprescr})
\bqa
\label{eq:exex1}
EEI= -8\int [d^4q_1] [d^4q_2] \frac{\hat \mu^2|_2 (\rlap/q_1 + \rlap/k_1)}{\qbar^4_1\qbar^2_2 \qbar^2_{12} (\qbar^2_1+k_1^2+2 (q_1 \cdot k_1))}.
\eqa   
Here, we mark the $\hat \mu^2|_2$ with a hat because {\em it is acting only on the $q_2$ sub-integral}. One finds\footnote{See appendix~\ref{app:exx}.}
\bqa
\label{eq:eei}
\!\!\!\!\!\!\!\!EEI(\mur)&=& \frac{2}{3} i\pi^2 \rlap/ k_1 \int [d^4q] 
\frac{1}{\qbar^2 (\qbar^2+k_1^2+2 (q\cdot k_1))} =
EEI_{b}+EEI_{a}(\mur)
\nl
\!\!\!\!\!\!\!\!&=&  V_0 \frac{4-n}{3}  \rlap/ k_1 \left\{ 
\int \frac{d^nq}{\mur^{-2 \epsilon}}
\frac{1}{q^2 (q^2+k_1^2+2 (q\cdot k_1))}
   -
\lim_{\mu \to 0}\int \frac{d^nq}{\mur^{-2 \epsilon}}\left.\frac{1}{\qbar^4}\right|_{\mu=\mur}\right\}
\eqa
where the last representation is suitable to be used to prove
eq.~(\ref{eq:fixcanc}). A further example of computation of an $EEI$ via sub-prescription is given in appendix~\ref{sec:exsub}.

Finally, it is important to realize that $EEI$s also arise in the intermediate steps of the calculation of the QCD correlation functions without external fermions discussed in section~\ref{sec-tloop}.
From eq.~(\ref{eq:compen}) and the universality\footnote{See section~\ref{sec:res}.} of the coupling constant extracted from all of the three-point vertices in figure~\ref{fig:1} we infer that one must find zero when summing up the $EEI$s from all contributing diagrams. Indeed, we have explicitly checked that this happens in the case of the $N_f$ corrections to the gluon and ghost propagators.
\section{Results} 
\label{sec:res}
The main result of our calculation is the list of QCD renormalization constants in appendix~\ref{sec:app0}. When setting $(\lambda,\delta)= (1,0)$ the known $\overline{\rm MS}$ 
formulae~\cite{Egorian:1978zx,Muta:2010xua,Mihaila:2012pz} are recovered.
The choice $(\lambda,\delta)= (0,1)$ corresponds to $\DRFDR$, i.e. the renormalization scheme which reproduces the FDR QCD Green's functions up to two loops.

As required by the QCD S-T identity
\bqa
\label{eq:3ratios}
\frac{Z_{GGG}}{Z_{GG}}= \frac{Z_{Gcc}}{Z_{cc}}= \frac{Z_{G\Psi\Psi}}{Z_{\Psi\Psi}}
= \sqrt{Z_{\alpha_S}}\sqrt{Z_{GG}},
\eqa
we have verified that the three ratios
\bqa
\label{eq:ratios}
\frac{1}{Z_{GG}} \left(\frac{Z_{GGG}}{Z_{GG}}\right)^2,~~~ 
\frac{1}{Z_{GG}} \left(\frac{Z_{Gcc}}{Z_{cc}}\right)^2,~~~
\frac{1}{Z_{GG}} \left(\frac{Z_{G\Psi\Psi}}{Z_{\Psi\Psi}}\right)^2,
\eqa
produce the same coupling constant renormalization, namely $Z_{\alpha_S}(\lambda,\delta)$ in eq.~(\ref{eq:zalpha}).
In the case of $Z_{\alpha_S}(0,1)$ this provides a stringent test on the universality of FDR.

Using $m_q \ne 0$ in $G_3^\ell$ gives the quark mass renormalization constant 
in  eq.~(\ref{eq:zm}). Again the correct $\overline{\rm MS}$ result~\cite{Muta:2010xua} is reproduced with $\lambda=1$ and $\delta=0$.
As an extra check, we have verified that any $m_q$ dependence drops 
in $Z_{\Psi \Psi}$, as should be. 

Eq.~(\ref{eq:zalpha}) can be used to infer the coupling constant shift between
FDR and $\overline{\rm MS}$
\bqa
\label{eq:shal}
 \frac{Z_{\alpha_S}(1,0)}{Z_{\alpha_S}(0,1)}=
    \frac{\alpha_S^{\text{FDR}}}{\alpha_S^{\overline{\text{MS}}}}
    &=& 
        1 +  \left(\frac{\alpha_S^{\overline{\text{MS}}}}{4\pi}\right)\frac{N_c}{3}
      +  \left(\frac{\alpha_S^{\overline{\text{MS}}}}{4\pi}\right)^2
        \Bigg\{
            \frac{89}{18} N_c^2 + 8 N_c^2 f 
 \nl
&&             +N_f
            \left[
                N_c - \frac{3}{2}C_F - 
                f 
                \left(
                    \frac{2}{3}N_c + \frac{4}{3}C_F
                \right)
            \right]
        \Bigg\}.
\eqa
In an analogous way eq.~(\ref{eq:zm}) produces the quark mass shift
\bqa
\label{eq:shzm}
    \frac{m_q^{\text{FDR}}}{m_q^{\overline{\text{MS}}}} &=& 
        1 - C_F
        \left(\frac{\alpha_S^{\overline{\text{MS}}}}{4\pi}\right) 
        + C_F
        \left(\frac{\alpha_S^{\overline{\text{MS}}}}{4\pi}\right)^2
        \Bigg\{
            \frac{77}{24} N_c - \frac{5}{8}C_F + 
            f\left(9 N_c + \frac{11}{3} C_F \right) \nl
&&+ N_f \left(\frac{1}{4} - \frac{2}{3} f\right)
        \Bigg\}.
\eqa
Eqs.~(\ref{eq:shal}) and (\ref{eq:shzm}) provide the transition rules from
IR finite QCD quantities computed in FDR and their analogue in $\overline{\rm MS}$.

Finally, it is well known that one can use the coupling constant shift between two schemes at a given
order to relate the two beta functions  at one order
higher. In the following we sketch out the two-loop proof of this~\cite{Bern:2002zk} and use it to derive the three-loop beta function in FDR.
Let us consider two different renormalization schemes defined in terms
of two coupling constants $\alpha_A$ and $\alpha_B$ related
by a shift
\begin{equation}
   \label{eq:GeneralCouplingShift}
    \frac{\alpha_A}{4 \pi} 
    = \frac{\alpha_B}{4 \pi}
      \left[
          1 + c_1 \left( \frac{\alpha_B}{4 \pi}\right) 
            + c_2 \left(\frac{\alpha_B}{4 \pi}\right)^2 
            + O\left( \alpha_B^3\right)
      \right].
\end{equation}
The two beta functions 
\begin{equation}
    \beta_{A,B} = \mu \frac{d}{d\mu} \frac{\alpha_{A,B}}{4 \pi} = 
    \left(\frac{\alpha_{A,B}}{4 \pi}\right)^2
      \left[ 
          b_0^{A,B}  + b_1^{A,B} \left( \frac{\alpha_{A,B}}{4 \pi} \right)
        + b_2^{A,B} \left(\frac{\alpha_{A,B}}{4 \pi}\right)^2
                 + O\left( \alpha_{A,B}^3\right)
      \right]
\end{equation}
are linked by the chain rule
\begin{equation}
    \beta_A = \mu \frac{d }{d \mu} \frac{\alpha_A}{4 \pi} 
            = \mu \left( \frac{d }{d \mu} \frac{\alpha_B}{4 \pi} \right)
                              \frac{d \alpha_A}{d \alpha_B}
            = \beta_B \frac{d \alpha_A}{d \alpha_B}.
\end{equation}
Writing both sides in terms of the same $\alpha$ through eq.~(\ref{eq:GeneralCouplingShift}) we find the standard result of scheme
independence up to two loops, i.e. $b_0^A = b_0^B$ and $b_1^A = b_1^B$ as
well as an expression for the three-loop beta function in scheme $B$,
using only two-loop information from scheme $A$
\begin{equation}
    b_2^B = b_2^A + c_1 b_1^A + \left( c_1^2 - c_2 \right) b_0^A.
\end{equation}
To calculate the three-loop beta function in FDR we 
use the three-loop $\overline{\rm MS}$ beta function in~\cite{Larin:1993tp} together with the values from from eq.~(\ref{eq:shal}). This gives 
\bqa
b_2^{\text{FDR}} &=& N_c ^ 3 \left( -\frac{3610}{27}  -\frac{176}{3} f \right) 
+ N_f ^ 2 \left(
   -\frac{40}{9} C_F - \frac{43}{27} N_c + f \left( -\frac{16}{9} C_F -
   \frac{8}{9} N_c \right) \right)
\nl
&&+ N_f \left( \frac{1331}{27} N_c ^ 2 +
   \frac{292}{9} N_c C_F - 2 C_F ^ 2 + f \left( \frac{140}{9} N_c ^ 2 +
   \frac{88}{9} N_c C_F \right) \right). 
\eqa

\section{Fixing the UV behaviour of FDH}
\label{sec:fdh}
Recently it has been observed that the original formulation of the FDH method breaks unitarity~\cite{Kilgore:2011ta}. This was seen as the standard renormalization program fails to remove all UV poles at high enough perturbative orders. 
During our work, it proved
of interest to try and investigate this problem from our renormalization
scheme perspective. Using the technology developed in the previous
sections we can easily perform an analysis of the validity of the
interpretation of FDH as a perturbative description of QFT.

We begin by setting up the problem as in eqs.~(\ref{eq:fond}).
We first recall eq.~(\ref{eq:fond1}) for an arbitrarily renormalized, dimensionally regulated correlation function. Next, we equate it to correlation functions calculated in the $\overline{\text{FDH}}$ scheme up to two loops, i.e. 
\begin{subequations}
\begin{align}
G^{\ell}_{\text{renormalized}}&=G^{\ell}_{\text{bare}} + (\ell\text{-loop-CTs}) +
\ldots + (1\text{-loop-CTs}),
\label{eq:6.1}
\\
G^{1}_{\text{renormalized}} &= 
    \overline{\text{MS}}
    \left\{ 
        G^1_{\text{bare}}(n_s=4)
    \right\},
\label{eq:6.1a}
\\
G^{2}_{\text{renormalized}} &= 
    \overline{\text{MS}}
    \left\{ 
        G^2_{\text{bare}}(n_s=4) + (1\text{-loop-CTs})|_{n_s=4}
    \right\}.
\label{eq:6.1b}
\end{align}
\end{subequations}
The Green's functions in the r.h.s. of eqs.~(\ref{eq:6.1a}) and (\ref{eq:6.1b}) are calculated in the $\overline{\text{FDH}}$ scheme by setting $n_s=4$
as dictated\footnote{Here and in the following we explicitly write
the values of $n_s$ to highlight if contributions are regulated in FDH ($n_s=4$) or DReg ($n_s=n$). In a fully general FDH scheme $n_s$ is an arbitrary parameter, but we restrict to the $n_s= 4$ case.}
and subtracting both poles and universal constants through the
$\overline{\text{MS}}\{x\}$ operation.

If $\overline{\text{FDH}}$ amounts
to a different renormalization scheme to $\overline{\text{MS}}$, then
the $\ell\text{-loop-CTs}$ should satisfy the renormalizability
condition~(\ref{eq:rencon}), i.e. they should be proportional to the Born and
local. We can proceed to calculate these at each order in a similar
way to section~\ref{sec:calculation}, making use of information from the FDR
vacuum part. At one loop, the calculation is simple and provides the
exact same $Z^1$ as FDR, indicating that $\overline{\text{FDH}}$
exhibits the same coupling shift from $\overline{\text{MS}}$. At two
loops the computation is more complicated but in general we can write the
following expression for what {\em should} be the two-loop counterterms:
\bqa
\begin{tabular}{rrllll}
 $(2\text{-loop-CTs}) =$ & $\!\!\!\!\overline{\text{MS}}\,\{$ & 
         $\hspace{-0.43cm}GV$ & $\hspace{-0.4cm}+\,SV$  & $\hspace{-0.4cm}-\,EEI$ & $\hspace{-0.4cm}+\,CT\}|_{n_s= 4}$  \\
                         & $\!\!\!\!$                 $-\,[\,$  & 
         $\hspace{-0.43cm}GV$ & $\hspace{-0.4cm}+\,SV$  & $\hspace{-0.4cm}-\,EEI$ & $\hspace{-0.4cm}+\,CT]\,|_{n_s= n}$. 
\end{tabular}
\eqa
We now split $SV$ and $EEI$ as in eqs.~(\ref{eq:subpiecep})
and~(\ref{eq:spliteei}).
As $n_s=n$ in the second line, we can make use of eq.~(\ref{eq:fixcanc}).
A direct calculation of the first line gives instead a
striking different result for $n_s=4$: the $CT$ part cancels the $b$ piece
of the sub-vacuum, but {\it not} the $EEI$. In summary
\bqa
\label{eq:fdhfix}
\begin{tabular}{rrllll}
 $\!\!(2\text{-loop-CTs}) =$ & $\!\!\!\!\overline{\text{MS}}\,\{$ & 
         $\hspace{-0.43cm}GV$ & $\hspace{-0.4cm}+\,SV^\prime_a$  & $\hspace{-0.4cm}-\,EEI_a$ & $\hspace{-1.47cm}-\,EEI_b\}|_{n_s= 4}$  \\
    $\!\!$                     & $\!\!\!\!$                 $-\,[\,$  & 
         $\hspace{-0.43cm}GV$ & $\hspace{-0.4cm}+\,SV^\prime_a$  & $\hspace{-0.4cm}-\,EEI_a]|_{n_s= n}$. &  
\end{tabular}
\eqa
This is problematic for FDH as the $EEI_{b}$ term does not in general
satisfy the renormalizability condition. That is, we see that for
$EEI_{b} \ne 0$, $\overline{\text{FDH}}$ cannot be interpreted as
a perturbative description of QFT. Nevertheless, as we have seen
previously, the $EEI_{b}$ contribution is zero for external gauge states
and so the only expected problems come with external fermions, just as
experienced in~\cite{Kilgore:2011ta}.

The form of equation~(\ref{eq:fdhfix}) naturally suggests
a new definition of FDH which satisfies the renormalizability condition.
Let us consider changing the bare two-loop FDH correlation function in the following way\footnote{It is important to understand that the change we make in equation 
\eqref{eq:FDHRedefinition} can be read directly from the diagrams {\em without
the computation of counterterms.}} 
\begin{equation}
    \label{eq:FDHRedefinition}
    G^2_{\text{bare}}|_{n_s=4} \rightarrow 
    G^2_{\text{bare}}|_{n_s=4} + EEI_{b}|_{n_s=4}.
\end{equation}
This now allows us to write down a working FDH analogue of
eq.~(\ref{eq:z2fdr})
\bqa
\begin{tabular}{rrllll}
 $\!\!Z^{2,FDH'}G^0 =$ & $\!\!\!\!\overline{\text{MS}}\,\{$ & 
         $\hspace{-0.43cm}GV$ & $\hspace{-0.4cm}+\,SV^\prime_a$  & $\hspace{-0.4cm}-\,EEI_a\}|_{n_s= 4}$ &    \\
    $\!\!$                     & $\!\!\!\!$                 $-\,[\,$  & 
         $\hspace{-0.43cm}GV$ & $\hspace{-0.4cm}+\,SV^\prime_a$  & $\hspace{-0.4cm}-\,EEI_a]\,|_{n_s= n}$. &  
\end{tabular}
\eqa
Here we call this modified definition $\FDRFDH$. Its
renormalization constants $Z^{\ell,\text{FDH}'}$ are given
in appendix~\ref{sec:app0} and correspond to the case $(\lambda,\delta)= (0,0)$.
From these we are able to calculate the analogous version of eq.~(\ref{eq:3ratios}) in this scheme, indicating that the QCD S-T identity is respected,
{\em even with external fermions}\footnote{This is the important new case
as the $EEI$s are zero for the correlation functions with external
gauge states.}. What's more, we verify that in all cases the coupling
constant shift agrees with the literature value~\cite{Bern:2002zk}, 
i.e. $\FDRFDH$ is equivalent to FDH when the latter scheme makes sense and provides consistent predictions in all the other cases. 
Together these results suggest that, at
least off-shell, this definition does not face the renormalization
difficulties of the original FDH formulation. Furthermore, as we newly
have control over the fermion sector, we are able to calculate a mass
renormalization constant, and thereby a mass shift between
$\overline{\text{MS}}$ and $\FDRFDH$
\begin{equation}
    \frac{m_q^{\text{FDH}'}}{m_q^{\overline{\text{MS}}}} = 
        1 - 
        C_F\left(\frac{\alpha_S^{\overline{\text{MS}}}}{4\pi}\right) +
        C_F
        \left(\frac{\alpha_S^{\overline{\text{MS}}}}{4\pi}\right)^2
        \left\{
            \frac{29}{12} N_c - \frac{13}{2}C_F + 
            \frac{1}{4} N_f 
        \right\}.
\end{equation}

Our new definition offers a different perspective than the recently
proposed approaches to the unitarity based difficulties of FDH. Roughly
speaking, any solution requires a connection
between internal and external states. In the approach of~\cite{Broggio:2015ata}, evanescent operators are introduced, as in dimensional reduction, to make the external states behave like the internal ones, thereby introducing a series of evanescent couplings.
Conversely, our approach based upon sub-integration consistency
makes internal states behave like external ones, and so does not
require the introduction of any new fields whilst maintaining $n_s=4$
spin degrees of freedom.
 
\section{Summary and outlook} 
As the FDR UV subtraction is consistently encoded in the definition of a four-dimensional and finite loop integration, the FDR approach to QFT does not require the introduction of counterterms in the Lagrangian. In particular, an order-by-order renormalization is avoided: the $\ell^{th}$ perturbative order is computed by only looking at $\ell$-loop Feynman diagrams.

We have proven, up to two loops in QCD, that FDR is equivalent to a particular renormalization scheme of Dimensional Regularization, dubbed $\DRFDR$, whose 
renormalization constants, extracted from two- and three-point vertices, obey the Slavnov-Taylor identities. $\DRFDR$ and $\overline{\rm MS}$ are related by shifts in $\alpha_s$ and $m_q$ that we have explicitly computed. The transition rules derived in this paper can be used to translate calculations of IR finite quantities from FDR to $\overline{\rm MS}$.

During our analysis, we have identified a bottom-up interpretation of the failure of the naive FDH scheme beyond one loop. FDR provides a natural fix to this: it automatically generates the finite pieces needed to restore the renormalizability of FDH. These extra terms are computed without introducing 
$\epsilon$-scalars nor evanescent quantities in the Lagrangian and can be directly read off from two-loop diagrams.
Including such FDR inspired terms in FDH defines a consistent renormalizable scheme that we have called $\FDRFDH$.

In this paper we have concentrated our focus on IR finite quantities. The possibility of consistently using FDR to regulate final state soft/collinear divergences at one-loop has been proven in ref.~\cite{Pittau:2013qla}. The study of FDR in the presence of NNLO IR singularities is left for future investigations.

\acknowledgments
This research was supported by the European Commission through contracts ERC-2011-AdG No 291377 (LHCtheory) and PITN-GA-2012-316704 (HIGGSTOOLS). We also thank
the project FPA2013-47836-C3-1-P.
Our figures are prepared with {\tt Axodraw}~\cite{Vermaseren:1994je}.

\appendix

\section{The renormalization constants of the QCD vertices}
\label{sec:app0}
In the following we list the renormalization constants of the QCD correlation functions of figure~\ref{fig:1} up to two loops in the Feynman--'t Hooft gauge
\bqa
Z_i(\lambda,\delta)= 1
+\left( \frac{ \alpha_S}{4 \pi}\right)   Z^1_i(\lambda,\delta)
+\left( \frac{\alpha_S}{4 \pi} \right)^2 Z^2_i(\lambda,\delta). 
\eqa
The three renormalization schemes we are interested in are parametrized by the values of $\lambda$ and $\delta$ given in table~\ref{table:1}.
$N_c$ is the number of colours, $N_f$ the number of active fermions, 
$C_F= \frac{N^2_c-1}{2 N_c}$ and
$\alpha_S= \alpha_S(\lambda,\delta)$ is the QCD coupling 
constant of each scheme.
The constant
\bqa
\label{eq:f}
f = \frac{i}{\sqrt{3}}\left( 
 {\rm Li}_2(e^{i\frac{\pi}{3}})
-{\rm Li}_2(e^{-i\frac{\pi}{3}})
\right)= -1.17195361\ldots
\eqa
originates from the evaluation of the two-loop global vacuum in DReg,
\begin{table}[h]
\begin{center} 
\begin{tabular}{l|c|c|c|}
          & $\DRFDR$ & $\overline{\rm MS}$ & $\FDRFDH$ \\ \hline
$\lambda$ &  0 & 1 & 0 \\ \hline
$\delta$  &  1 & 0 & 0 \\ \hline
\end{tabular}
\caption{$\lambda$ and $\delta$ in 
$\DRFDR$, $\overline{\rm MS}$ and $\FDRFDH$.
 }
\label{table:1}
\end{center}
\end{table}

%
%
%
%
\bqa
    Z_{GG}&=& Z_1(\lambda,\delta)
\hskip 0.78\textwidth
\nl
&=& 1 + 
\frac{ \alpha_S }{4 \pi} 
\left\{
\frac{1}{ \epsilon } 
\left[ 
    \frac{5}{3} N_c - \frac{2}{3} N_f
\right] 
+ 
\frac{1}{3} N_c (1 - \lambda)
\right\}
\nl
&&+ 
\left( \frac{\alpha_S}{4 \pi} \right)^2 
\bigg\{ 
    \frac{1}{\epsilon^2} 
           \left[ -\frac{25}{12} N_c ^ 2 + \frac{5}{6} N_c N_f\right] 
\nl
&&+\frac{1}{\epsilon} 
            \left[ \frac{65}{24} N_c ^ 2 + \frac{1}{6} \lambda N_c ^ 2 + N_f
   \left( -C_F - \frac{5}{4} N_c \right) \right] 
\nl
&&+
             (1 - \lambda) \frac{15}{8} N_c ^ 2 + \delta \left(
   \frac{9}{16} N_c ^ 2 + \frac{7}{2} N_c ^ 2 f \right)  
\nl
&&+
            N_f  \left( -(1 - \lambda) C_F + \delta \left( \frac{7}{8} N_c -
   \frac{1}{2} C_F + f \left( -\frac{4}{3} C_F - \frac{1}{3} N_c \right)
   \right) \right)\bigg\},
\eqa
%
%
\bqa
    Z_{cc} &=&  Z_2(\lambda,\delta)
\hskip 0.78\textwidth
\nl
&=& 1 + 
\frac{ \alpha_S }{4 \pi} 
\frac{1}{ \epsilon } \left[ 
    \frac{1}{2} N_c
\right]
\nl
&&+ 
\left( \frac{\alpha_S}{4 \pi} \right)^2 
\bigg\{ 
    \frac{1}{\epsilon^2} 
           \left[ -N_c ^ 2 + \frac{1}{4} N_c N_f\right] 
\nl
&&+\frac{1}{\epsilon} 
            \left[ \frac{37}{48} N_c ^ 2 + \frac{1}{4} \lambda N_c ^ 2 -
   \frac{5}{24} N_c N_f \right] 
\nl
&&+ (1 - \lambda)\frac{11}{48} N_c ^ 2 + \delta \left(
   \frac{19}{32} N_c ^ 2 + \frac{3}{2} N_c ^ 2 f \right) + \delta N_f \left(
   \frac{1}{16} N_c - \frac{1}{6} N_c f \right) \bigg\},
\eqa
%
%
\bqa
    Z_{\Psi\Psi} &=& Z_3(\lambda,\delta)
\hskip 0.78\textwidth
\nl
&=& 1 + 
\frac{ \alpha_S }{4 \pi} 
\left\{
\frac{1}{ \epsilon } 
\left[
    -C_F
\right] 
+ 
C_F (1 - \lambda)
\right\}
\nl
&&+ 
\left( \frac{\alpha_S}{4 \pi} \right)^2 
\bigg\{ 
    \frac{1}{\epsilon^2} 
           \left[ N_c C_F + \frac{1}{2} C_F ^ 2\right] 
\nl
&&+
  \frac{1}{\epsilon} 
            \left[ -\frac{1}{4} C_F ^ 2 - \frac{17}{4} N_c C_F + \lambda C_F
   ^ 2 + \frac{1}{2} N_f C_F \right]  
\nl
&&+
             (1 - \lambda)\left(\frac{13}{4} N_c C_F - \frac{1}{2} C_F ^ 2\right)
   + \delta \left( -\frac{1}{8} C_F ^ 2 - \frac{19}{8} N_c C_F + f \left(
   -\frac{1}{3} C_F ^ 2 - 4 N_c C_F \right) \right)  
\nl
&&+
            N_f  \left( -(1 - \lambda)\frac{3}{4} C_F + \delta \left( C_F +
   \frac{4}{3} C_F f \right) \right)\bigg\},
\eqa
%
%
%
%
\bqa
    Z_{GGG} &=& Z_4(\lambda,\delta)
\hskip 0.78\textwidth
\nl
&=& 1 + 
\frac{ \alpha_S }{4 \pi} 
\left\{
\frac{1}{ \epsilon } 
\left[
    \frac{2}{3} N_c - \frac{2}{3} N_f
\right] 
+ 
\frac{1}{3} N_c (1 - \lambda)
\right\}
\nl
&& + 
\left( \frac{\alpha_S}{4 \pi} \right)^2 
\bigg\{
    \frac{1}{\epsilon^2} 
           \left[ -\frac{13}{8} N_c ^ 2 + \frac{5}{4} N_c N_f\right] 
\nl
&&+
  \frac{1}{\epsilon} 
            \left[ \frac{59}{48} N_c ^ 2 + \frac{1}{4} \lambda N_c ^ 2 + N_f
   \left( -C_F - \frac{25}{24} N_c \right) \right] 
\nl
&& + 
             (1 - \lambda) \frac{79}{48} N_c ^ 2 + \delta \left(
   -\frac{13}{32} N_c ^ 2 + \frac{5}{4} N_c ^ 2 f \right)   
\nl
&&+
            N_f  \left( -(1 - \lambda)C_F + \delta \left( \frac{13}{16} N_c
   - \frac{1}{2} C_F + f \left( -\frac{4}{3} C_F - \frac{1}{6} N_c \right)
   \right) \right)\bigg\},
\eqa
%
%
\bqa
    Z_{Gcc} &=& Z_5(\lambda,\delta)
\hskip 0.78\textwidth
\nl
&=& 1 + 
\frac{ \alpha_S }{4 \pi} 
\frac{1}{ \epsilon } \left[ 
    -\frac{1}{2} N_c
\right]
\nl
&&+ 
\left( \frac{\alpha_S}{4 \pi} \right)^2 
\bigg\{ 
    \frac{1}{\epsilon^2} 
           \left[ \frac{5}{8} N_c ^ 2\right] +  \frac{1}{\epsilon} 
            \left[ -\frac{3}{8} N_c ^ 2 \right] 
 +           \delta \left( -\frac{3}{8} N_c ^ 2 - \frac{3}{4}
   N_c ^ 2 f \right)\bigg\},
\eqa
%
%
\bqa
    Z_{G\Psi\Psi} &=& Z_6(\lambda,\delta)
\hskip 0.78\textwidth
\nl
&=& 1 + 
\frac{ \alpha_S }{4 \pi} 
\left\{
\frac{1}{ \epsilon } 
\left[ 
    -C_F - N_c
\right] 
+ 
C_F (1 - \lambda)
\right\}
\nl
&&+ 
\left( \frac{\alpha_S}{4 \pi} \right)^2 
\bigg\{ 
    \frac{1}{\epsilon^2} 
           \left[ -1 + \frac{25}{8} N_c ^ 2 + \frac{1}{2} C_F ^ 2 -
   \frac{1}{4} N_c N_f\right] 
\nl
&&+
  \frac{1}{\epsilon} 
            \left[ \frac{21}{8} - \frac{1}{4} C_F ^ 2 - \frac{181}{48} N_c ^
   2 + \lambda \left( -\frac{1}{2} + \frac{1}{4} N_c ^ 2 + C_F ^ 2 \right) +
   N_f \left( \frac{5}{24} N_c + \frac{1}{2} C_F \right) \right]  
\nl
&&+
            (1 - \lambda) \left( -\frac{13}{8} + \frac{67}{48} N_c ^ 2 -
   \frac{1}{2} C_F ^ 2 
\right)  
\nl
&&+
            \delta 
    \left( \frac{19}{16} - \frac{1}{8} C_F ^ 2 -
   \frac{69}{32} N_c ^ 2 + f \left( 2 - \frac{1}{3} C_F ^ 2 - \frac{17}{4}
   N_c ^ 2 \right) \right) 
\nl
&&+
            N_f  \left( -(1 - \lambda)\frac{3}{4} C_F + \delta \left( C_F -
   \frac{1}{16} N_c + f \left( \frac{1}{6} N_c + \frac{4}{3} C_F \right)
   \right) \right)\bigg\}.
\eqa
The above results can be used to derive the renormalization constant of
the QCD coupling through eqs.~(\ref{eq:ratios})
%
%
\bqa
\label{eq:zalpha}
     Z_{\alpha_S}(\lambda,\delta) &=&
1 + 
\frac{ \alpha_S }{4 \pi} 
\left\{
\frac{1}{ \epsilon } 
\left[ 
    -\frac{11}{3} N_c + \frac{2}{3} N_f
\right] 
-\frac{1}{3} N_c (1 - \lambda)
\right\}
\nl
&&+ 
\left( \frac{\alpha_S}{4 \pi} \right)^2 
\bigg\{
    \frac{1}{\epsilon^2} 
           \left[ \frac{121}{9} N_c ^ 2 + \frac{4}{9} N_f ^ 2 - \frac{44}{9}
   N_c N_f\right] 
\nl
&&+
  \frac{1}{\epsilon} 
            \left[ -\frac{29}{9} N_c ^ 2 - \frac{22}{9} \lambda N_c ^ 2 +
   N_f \left( \frac{11}{9} N_c + C_F + \frac{4}{9} \lambda N_c \right)
   \right]  
\nl
&&+
             \left( -(1 - \lambda)\frac{20}{9} N_c ^ 2 + \delta \left(
   -\frac{5}{2} N_c ^ 2 - 8 N_c ^ 2 f \right) \right) 
\nl
&&+
            N_f  \left( (1 - \lambda) C_F + \delta \left( \frac{1}{2} C_F -
   N_c + f \left( \frac{2}{3} N_c + \frac{4}{3} C_F \right) \right)
   \right)\bigg\}.
\eqa
Finally, computing $G^\ell_{\Psi \Psi}$ with $m_q \ne 0$ gives the renormalization
constant associated with the quark mass
%
%
%
%
\bqa
\label{eq:zm}
    Z_{m_q} &=&
1 + 
\frac{ \alpha_S }{4 \pi} 
\left\{
\frac{1}{ \epsilon } 
\left[ 
    -3 C_F
\right] 
+ 
C_F (1 - \lambda)
\right\}
\nl
&&+ 
\left( \frac{\alpha_S}{4 \pi} \right)^2 
\bigg\{ 
    \frac{1}{\epsilon^2} 
           \left[ \frac{11}{2} N_c C_F + \frac{9}{2} C_F ^ 2 - N_f
   C_F\right]
\nl
&&
  +\frac{1}{\epsilon} 
            \left[ -\frac{15}{4} C_F ^ 2 - \frac{85}{12} N_c C_F + \lambda
   \left( 3 C_F ^ 2 - N_c C_F \right) + \frac{5}{6} N_f C_F \right]  
\nl
&&
  +           (1 - \lambda)\left( \frac{15}{2} C_F ^ 2 - \frac{11}{4} N_c C_F\right)
   - \delta \left(\frac{47}{8} C_F ^ 2 + \frac{19}{24} N_c C_F + f \left(
   \frac{11}{3} C_F ^ 2 + 9 N_c C_F \right) \right)   
\nl
&&
    +        N_f  \left( -(1 - \lambda)\frac{1}{4} C_F + \delta \frac{2}{3} 
   C_F f \right)\bigg\}.
\eqa

\section{A two-loop diagram}
\label{sec:ex}
As an example of the algorithm we use to extract and compute the FDR vacuum we work out in detail one among the two-loop diagrams contributing to $G^2_2$ in figure~\ref{fig:1}, namely the ghost-loop correction depicted 
in figure~\ref{fig:2}. 
\FIGURE{
\begin{picture}(190,83)(0,0)
\SetScale{1.5}
\SetWidth{0.5}
\SetOffset(-30,10)
\DashArrowLine(30,0)(60,0){2.5}
\DashArrowLine(60,0)(120,0){2.5}
\DashArrowLine(120,0)(150,0){2.5}
\SetWidth{0.2}
%
\LongArrow(32,7)(45,7)
\Text(58,15)[b]{$k_1$}
\LongArrowArcn(90,30)(20,156,114)
\Text(112,71)[r]{$q_2$}
\LongArrowArc(90,0)(37,145,165)
\Text(84,26)[r]{$q_1$}
\SetWidth{0.5}
\GlueArc(90,0)(30,0,60){2}{5}
\GlueArc(90,0)(30,120,180){2}{5}
\DashArrowArcn(90,30)(15,0,-180){2.5}
\DashArrowArcn(90,30)(15,-180,0){2.5}
\end{picture}
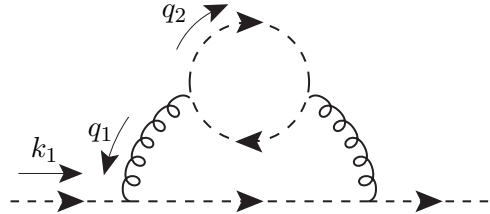
\caption{Two-loop diagram contributing to the ghost self-energy.}
\label{fig:2}
}
The integrand of the corresponding amplitude is proportional to
\bqa
\label{eq:diago}
J(q_1,q_2)= \frac{(k_1 \cdot q_{12}) (q_1+k_1) \cdot q_2}{q_1^4 D_1 q^2_2 q^2_{12}}, \nl
\eqa
with $q_{12}= q_1+q_2$ and ${D_1}= (q_1+k_1)^2$. 
To read off $J(q_1,q_2,\mu^2)$ from $J(q_1,q_2)$ 
we apply the shift of eq.~(\ref{eq:shift}) 
in {\em both} the numerator and the denominator
and simplify reducible numerators.
This gives
\bqa
\label{eq:integtot}
J(q_1,q_2,\mu^2) &=& \frac{(k_1 \cdot q_2)^2}{\qbar_1^4 \bar D_1 \qbar^2_2 \qbar^2_{12} }
 -\frac{k_1^2}{2}\frac{(k_1 \cdot q_2)}{\qbar_1^4 \bar D_1 \qbar^2_2 \qbar^2_{12} }
  +\frac{1}{2} \frac{(k_1 \cdot q_2)}{\qbar_1^4 \qbar^2_2 \qbar^2_{12}}
             -\frac{k_1^2}{4} \frac{1}{\qbar_1^4 \bar D_1 \qbar^2_2 }
   +\frac{1}{2} \frac{(k_1 \cdot q_2)}{\qbar_1^4 \bar D_1 \qbar^2_2 }
   +\frac{1}{4} \frac{1}{ \qbar_1^4 \qbar^2_2} \nl
          &&+\frac{k_1^2}{4} \frac{1}{\qbar_1^4 \bar D_1 \qbar^2_{12} } 
-\frac{1}{2} \frac{(k_1 \cdot q_2)}{\qbar_1^4 \bar D_1 \qbar^2_{12} } 
-\frac{1}{4} \frac{1}{\qbar_1^4 \qbar^2_{12}} 
          +\frac{k_1^2}{4} \frac{1}{\qbar_1^2 \bar D_1 \qbar^2_2 \qbar^2_{12}}
            -\frac{(k_1 \cdot q_2)}{\qbar_1^2 \bar D_1 \qbar^2_2 \qbar^2_{12}} \nl
&&-\frac{1}{4} \frac{1}{\qbar_1^2 \qbar^2_2 \qbar^2_{12}} 
-\frac{1}{4} \frac{1}{\qbar_1^2 \bar D_1\qbar^2_2}
+ \frac{1}{4} \frac{1}{\qbar_1^2\bar D_1 \qbar^2_{12}}
+ \frac{1}{4} \frac{1}{\qbar_2^2\bar D_1 \qbar^2_{12}}.
\eqa
Notice that the integrand manipulations we have performed 
so far are allowed both in FDR and DReg.
In the following we concentrate on the first tensor
\bqa
\label{eq:J0}
J_0(q_1,q_2,\mu^2) = \frac{q_2^\alpha q_2^\beta }{\qbar_1^4\bar D_1 \qbar^2_2 \qbar^2_{12} },
\eqa 
and explicitly derive the splitting 
\bqa
J_0(q_1,q_2,\mu^2) = \left[J_{0,{\rm INF}}(q_1,q_2,\mu^2)\right] +J_{0,{\rm F}}(q_1,q_2,\mu^2) 
\eqa
needed to define the FDR integral
\bqa
\int[d^4q_1] [d^4q_2]\, J_0(q_1,q_2,\mu^2) = \lim_{\mu \to 0} \int d^4q_1   d^4q_2 
\,J_{0,{\rm F}}(q_1,q_2,\mu^2). 
\eqa
The other terms in eq.~(\ref{eq:integtot}) are easier and can be treated analogously.

To analyze the UV behaviour of a two-loop integrand $J$ 
we introduce the four operators
$d_0(J)$, $d_1(J)$, $d_2(J)$ and $ d_{12}(J)$, which indicate how the integral over it behaves for large values of the integration momenta when $q_1 \to \infty$ and $q_2\to \infty$ independently ($d_0(J)$) or when $q_i$ is fixed ($d_i(J)$). Thus, UV divergences occur when $d_i(J) \ge 0$ for some $i$. In our case, as
\bqa
d_0(J_0)= 0,~~ d_1(J_0)= 2,~~ d_2(J_0) < 0,~~ d_{12}(J_0) < 0, 
\eqa
$J_0(q_1,q_2,\mu^2)$ has a logarithmic global UV divergence and is quadratically divergent in one of its sub-integrations.
We now apply twice the identity
\bqa
\label{eq:ext2}
\frac{1}{\qbar^2_{12}}= \frac{1}{\qbar^2_2}- \frac{q^2_1+2 (q_1 \cdot q_2)}{\qbar^2_2 \qbar^2_{12}}
\eqa
and rewrite
\bqa
 J_0(q_1,q_2,\mu^2) &=& \left[J_1(q_1,q_2,\mu^2)\right]-\left[J_2(q_1,q_2,\mu^2)\right]+J_3(q_1,q_2,\mu^2),
\eqa
with
\bqa
\left[ J_1(q_1,q_2,\mu^2)\right] &=& \frac{1}{\qbar_1^4\bar D_1} \left[\frac{q_2^\alpha q_2^\beta}{\qbar^4_2}\right],\\
\label{eq:j2}
\left[ J_2(q_1,q_2,\mu^2) \right]&=& \frac{q^2_1}{\qbar_1^4\bar D_1}\left[\frac{q_2^\alpha q_2^\beta}{\qbar^6_2}\right]
+2 \frac{q_{1\gamma}}{\qbar_1^4\bar D_1}\left[\frac{q_2^\alpha q_2^\beta q_2^\gamma}{\qbar^6_2}\right],\\
J_3(q_1,q_2,\mu^2)&=& \frac{q_2^\alpha q_2^\beta(q^2_1+2 (q_1 \cdot q_2))^2}{\qbar_1^4\bar D_1\qbar^6_2 \qbar^2_{12}}.
\eqa
We see that $\left[J_1(q_1,q_2,\mu^2)\right]$ and $\left[J_2(q_1,q_2,\mu^2)\right]$ are factorizable integrands in which the 
$q_2 \to \infty$ behaviour is fully parametrized in terms of divergent integrands depending only on $\mu^2$.  Therefore, they belong to $\left[J_{0,{\rm INF}}(q_1,q_2,\mu^2) \right]$. In addition, since
\bqa
d_0(J_3)= 0,~~ d_1(J_3)= 0,~~ d_2(J_3) = 0,~~ d_{12}(J_3) < 0, 
\eqa
further infinities need to be extracted from it, that is achieved by rewriting
\bqa
\frac{1}{\bar D_1}= \frac{1}{\qbar_1^2}-\frac{p^2+2(p \cdot q_1)}{\qbar_1^2\bar D_1},
\eqa
which gives
\bqa
J_3(q_1,q_2,\mu^2)&=& \left[J_4(q_1,q_2,\mu^2)\right]-J_5(q_1,q_2,\mu^2)-J_6(q_1,q_2,\mu^2),
\eqa
where
\bqa
\label{eq:exglo}
\left[ J_4(q_1,q_2,\mu^2)\right] &=& \left[\frac{q_2^\alpha q_2^\beta(q^2_1+2 (q_1 \cdot q_2))^2}{\qbar_1^6 \qbar^6_2 \qbar^2_{12}}\right], \nl
J_5(q_1,q_2,\mu^2) &=& 4 \frac{q_2^\alpha q_2^\beta (q_1 \cdot q_2)^2(p^2+2(p \cdot q_1))}{\qbar_1^6\bar D_1\qbar^6_2 \qbar^2_{12}},\nl
J_6(q_1,q_2,\mu^2) &=& \frac{q_2^\alpha q_2^\beta(q^4_1+4 q^2_1 (q_1 \cdot q_2))(p^2+2(p \cdot q_1))}{\qbar_1^6\bar D_1\qbar^6_2 \qbar^2_{12}}.
\eqa
Thus, the non-factorizable integrand $\left[ J_4(q_1,q_2,\mu^2)\right]$ contributes to $\left[J_{0,{\rm INF}}(q_1,q_2,\mu^2)\right]$ and $J_6(q_1,q_2,\mu^2)$ is UV convergent.  
Furthermore, since
\bqa
d_0(J_5) < 0,~~ d_1(J_5)= 0,~~ d_2(J_5) < 0 ,~~ d_{12}(J_5) < 0, 
\eqa
a logarithmic sub-divergence is still present in $J_5(q_1,q_2,\mu^2)$, that gets separated when applying once more eq.~(\ref{eq:ext2})
\bqa
J_5(q_1,q_2,\mu^2)= \left[ J_7(q_1,q_2,\mu^2)\right]-J_8(q_1,q_2,\mu^2),
\eqa  
with
\bqa
\label{eq:j7}
\left[ J_7(q_1,q_2,\mu^2)\right] &=& 4 \frac{q_{1\gamma} q_{1\delta}(p^2+2(p \cdot q_1))}{\qbar_1^6\bar D_1} \left[\frac{q_2^\alpha q_2^\beta q_2^\gamma q_2^\delta}{\qbar^8_2}\right],\\
J_8(q_1,q_2,\mu^2) &=& 4 \frac{q_2^\alpha q_2^\beta (q_1 \cdot q_2)^2(p^2+2(p \cdot q_1))(q^2_1+2 (q_1 \cdot q_2))}{\qbar_1^6\bar D_1\qbar^8_2 \qbar^2_{12}}.
\eqa
In summary, $J_0(q_1,q_2,\mu^2)$ should be split as follows:
\bqa
\begin{tabular}{rll}
$J_{0,{\rm F}}(q_1,q_2,\mu^2)$ & $\!\!\!\!=$& $\!\!\!\!\!J_8(q_1,q_2,\mu^2)-J_6(q_1,q_2,\mu^2)$, \nl \nl
$\left[J_{0,{\rm INF}}(q_1,q_2,\mu^2) \right]$ & $\!\!\!\!=\!\!\!$ &
   $\!\!\!\!\!\left[ J_4(q_1,q_2,\mu^2) \right]
   +\left[ J_1(q_1,q_2,\mu^2) \right]  
   -\left[ J_2(q_1,q_2,\mu^2) \right]
   -\left[ J_7(q_1,q_2,\mu^2) \right]$. 
\end{tabular}
 \eqa
In the rest of the appendix we analyze the four terms contributing to 
$\left[J_{0,{\rm INF}}(q_1,q_2,\mu^2) \right]$ to establish their connection with 
$GV_2(\mu)$ and $SV_2(\mu)$ in eqs.~(\ref{eq:globv})--(\ref{eq:jsplit2}). 
\begin{itemize}
\item $\left[ J_4(q_1,q_2,\mu^2) \right]$: 

It gives a contribution to $GV_2(\mu)$. By using integration-by-parts one finds, for any value of $n$,
\bqa
&&\!\!\!\!\!\!\!\!\lim_{\mu\to 0}\int \frac{d^n q_1}{\mur^{-2 \epsilon}} \frac{d^n q_2}{\mur^{-2 \epsilon}}
\left[ J_4(q_1,q_2,\mu^2)\right]\nl
&&\!\!\!\!= g^{\alpha \beta}
\lim_{\mu\to 0}\int \frac{d^n q_1}{\mur^{-2 \epsilon}} \frac{d^n q_2}{\mur^{-2 \epsilon}} \left\{\left( \frac{4}{3n} - \frac{1}{4} \right)\left[\frac{1}{\qbar^4_1 \qbar^4_2}\right]+\left(\frac{1}{6}- \frac{4}{3n}  \right)
           \left[\frac{1}{\qbar^4_1 \qbar^2_2 \qbar^2_{12}}\right]\right\}.
\eqa
\item $\left[ J_1(q_1,q_2,\mu^2)\right]$:

Since it is proportional to $\mu^2$, it vanishes in the limit $\mu^2 \to 0$ we are interested in.
\item $\left[ J_2(q_1,q_2,\mu^2)\right]$:

The the odd-rank tensor in the r.h.s. of eq.~(\ref{eq:j2}) gives zero upon integration. On the other hand $\mu^2 \to 0$ is allowed in the $q_1$ integrand multiplying the rank-two tensor. Thus
\bqa
&&\lim_{\mu\to 0}\int \frac{d^n q_1}{\mur^{-2 \epsilon}} \frac{d^n q_2}{\mur^{-2 \epsilon}}
\left[ J_2(q_1,q_2,\mu^2)\right] \nl
&&~~= \frac{g^{\alpha \beta}}{4}
\lim_{\mu\to 0}\int \frac{d^n q_1}{\mur^{-2 \epsilon}} \frac{d^n q_2}{\mur^{-2 \epsilon}}
\left[\frac{1}{\qbar_2^4}\right] \left\{\left(\frac{1}{q_1^2 D_1}-
\left[\frac{1}{\qbar_1^4}\right] \right) + \left[\frac{1}{\qbar_1^4}\right] \right\}.
\eqa
The first term in the curly bracket contributes to $SV_2(\mu)$ with 
$A^\prime_2= -A_2= {g^{\alpha \beta}}/{4}$, while the second to 
$GV_2(\mu)$ with $F_{22}= 0$ and $F_{21}= {g^{\alpha \beta}}/{4}$. 

\item $\left[ J_7(q_1,q_2,\mu^2) \right]$: 

The integrand depending on $q_1$ in eq.~(\ref{eq:j7}) behaves as $1/q^4_1$ when $q_1 \to 0$. To disentangle the term which develops a $\ln \mu^2$ we rewrite\footnote{As mentioned in section~\ref{sec:calculation}, this separation is in general achieved by using the propagator identity in eq.~(\ref{eq:invprop}).}
\bqa
\frac{q_{1\gamma} q_{1\delta}(p^2+2(p \cdot q_1))}{\qbar_1^6\bar D_1}= 
 \frac{q_{1\gamma} q_{1\delta}}{\qbar_1^6}
-\frac{q_{1\gamma} q_{1\delta}}{\qbar_1^4\bar D_1}=
 \frac{q_{1\gamma} q_{1\delta}}{\qbar_1^6}
-\frac{q_{1\gamma} q_{1\delta}}{q_1^4 D_1},
\eqa
which results in
\bqa
\left[ J_7(q_1,q_2,\mu^2)\right] &=& 4 \left(
  \left[ \frac{q_{1\gamma} q_{1\delta}}{\qbar_1^6} \right]
- \frac{q_{1\gamma} q_{1\delta}}{q_1^4 D_1}   \right)
\left[\frac{q_2^\alpha q_2^\beta q_2^\gamma q_2^\delta}{\qbar^8_2}\right].
\eqa
Since the integral over $q_1$ is UV finite
\bqa
\lim_{\mu\to 0}\int \frac{d^n q_1}{\mur^{-2 \epsilon}} \frac{d^n q_2}{\mur^{-2 \epsilon}}
\left[ J_7(q_1,q_2,\mu^2)\right]
\eqa
only contributes to $SV_2(\mu)$. By using  tensor reduction and integration-by-parts identities one finds $A^\prime_2= -\frac{1}{24} g^{\mu \nu} + \frac{n-4}{24}p^\mu p^\nu$ and $A_2=\frac{1}{24}g^{\mu \nu}$ for this term.
\end{itemize}

\section{Sub-prescription example}
\label{sec:exsub}
Here we discuss a further example of the sub-prescription in order
to aid the understanding of the reader for future FDR calculations.
Consider the FDR integral
\bqa
   \int \FDRMeasure{q_1} \FDRMeasure{q_2}
          \frac{
\gamma_\beta (\feynSlash{q}_2 + \feynSlash{k_1}) \gamma_\alpha
(\feynSlash{q}_{12} + \feynSlash{k_1}) \gamma^\beta 
(\feynSlash{q}_{1} + \feynSlash{k_1}) \gamma^\alpha
}{\FDRbar{q}_1^2 \FDRbar{q}_2^2 \FDRbar{D}_1 \FDRbar{D}_2
          \FDRbar{D}_{12}},
\eqa
corresponding to the contribution to $G^2_3$ in figure~\ref{fig:fig3}, where $\FDRbar{D}_i = \FDRbar{q}^2_i + k_1^2 + 2 (q_i \cdot k_1)$.
\FIGURE{
\begin{picture}(200,115) (0,-15)
 \SetOffset(22,0)
    \SetWidth{1}
    \SetScale{0.8}
    \SetColor{Black}
    \GluonArc(77,50)(46,0,180) {6} {12}
    \Line[arrow,arrowpos=1.0,arrowlength=4,arrowwidth=2,arrowinset=0.2](87,107)(67,107)
    \Text(58.4,88)[lb]{$q_1$}
    \GluonArc(123,50)(46,180,0) {6} {12}
    \Line[arrow,arrowpos=1.0,arrowlength=4,arrowwidth=2,arrowinset=0.2](133,-6)(113,-6)
    \Text(95.2,-14.4)[lb]{$q_2$}

    \Line[arrow,arrowpos=0.45,arrowlength=5,arrowwidth=2,arrowinset=0.2](0,50)(31,50)
    \Text(6,45.6)[lb]{$k_1$}
    \Line[arrow,arrowpos=0.5,arrowlength=5,arrowwidth=2,arrowinset=0.2](31,50)(77,50)
    \Text(34.4,45.6)[lb]{$q_1\!+\!k_1$}
    \Line[arrow,arrowpos=0.5,arrowlength=5,arrowwidth=2,arrowinset=0.2](77,50)(123,50)
    \Text(72,28.8)[lb]{$q_{12}\!+\!k_1$}
    \Line[arrow,arrowpos=0.5,arrowlength=5,arrowwidth=2,arrowinset=0.2](123,50)(169,50)
    \Text(112.2,45.6)[lb]{$q_2\!+\!k_1$}
    \Line[arrow,arrowpos=0.55,arrowlength=5,arrowwidth=2,arrowinset=0.2](169,50)(200,50)
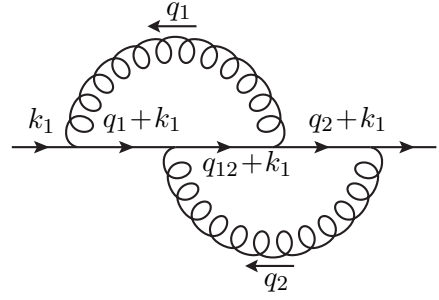
\caption{Diagram contributing to $G^2_3$. The fermion momenta follows the fermion line.}
\end{picture}
\label{fig:fig3}
}
We wish to discuss
how to extract the $EEI$ resulting from the sub-prescription, so as in
figure~\ref{fig:loop1} we consider the un-promoted numerator in order to find
the relevant terms.
In this diagram we have two sub-divergences, one for fixed $q_1$ and
another fixed $q_2$. The terms from the sub-prescription of each can be
extracted considering the sub-divergences independently. First we shall
consider $q_1$ fixed. Let us disconnect the divergent sub-diagram, as in
figure~\ref{fig:fig4}.
\FIGURE{
  \begin{picture}(200,100) (0,-15)
    \SetOffset(25,-17)
    \SetScale{0.8}
    \SetWidth{1}
    \SetColor{Black}

    \GluonArc(77,50)(46,0,180) {6} {12}
     \Text(10,45.6)[lb]{$\hat{\alpha}$}
    \Text(106.4,45.6)[lb]{$\alpha$}

    \GluonArc(123,50)(46,180,0) {6} {12}
     \Text(71.2,25)[lb]{$\beta$}
    \Text(144,25)[lb]{$\beta$}

    \Line[arrow,arrowpos=0.45,arrowlength=5,arrowwidth=2,arrowinset=0.2](0,50)(31,50)
    \Line[arrow,arrowpos=0.5,arrowlength=5,arrowwidth=2,arrowinset=0.2](31,50)(45,50)
    \Line[arrow,arrowpos=0.5,arrowlength=5,arrowwidth=2,arrowinset=0.2](60,50)(77,50)
    \Line[arrow,arrowpos=0.5,arrowlength=5,arrowwidth=2,arrowinset=0.2](77,50)(123,50)
    \Line[arrow,arrowpos=0.5,arrowlength=5,arrowwidth=2,arrowinset=0.2](123,50)(169,50)
    \Line[arrow,arrowpos=0.55,arrowlength=5,arrowwidth=2,arrowinset=0.2](169,50)(200,50)

    \CBox(67,85)(87,110){White}{White}
    \SetWidth{1.5}
    \put(59.2,68)
    {
        \Line(0,0)(0,30)
        \Line(6,0)(6,30)
    }

    \put(40,27.2)
    {
        \Line(0,0)(0,30)
        \Line(6,0)(6,30)
    }
  \end{picture}
\caption{The same diagram of figure~\ref{fig:fig3} from the point of view of the $q_2$ sub-integration.} 
\label{fig:fig4}
}
The numerator with its appropriate hatting reads
\begin{equation}
    N = \gamma_\beta (\feynSlash{q}_2 + \feynSlash{k_1}) \gamma_{{\alpha}}
        (\feynSlash{q}_{12} + \feynSlash{k_1}) \gamma^\beta 
        (\feynSlash{q}_{1} + \feynSlash{k_1}) \gamma^{\hat{\alpha}}
\end{equation}
and the numerator terms which give logarithmic divergences
in $q_2$ are
\bqa
    N^{(2)} 
        &=& \gamma_\beta \feynSlash{q}_2 \gamma_\alpha \feynSlash{q}_{2} 
          \gamma^\beta (\feynSlash{q}_{1} + \feynSlash{k_1}) \gamma^{\hat \alpha} \nl
      &=& -4 q^2_2 (\feynSlash{q}_{1} + \feynSlash{k_1})
    -4 \feynSlash{q}_2 (\feynSlash{q}_{1} + \feynSlash{k_1})\feynSlash{\hat q}_2.\nl
\eqa
The sub-prescription gives
\bqa
 N^{(2)} \to N^{(2)} +4 \mu^2|_2(\feynSlash{q}_{1} + \feynSlash{k_1}), 
\eqa
while the global prescription requires no change in $N^{(2)}$:
\bqa
 N^{(2)} \to N^{(2)}.
\eqa
Thus, we find the resulting contribution by subtracting the (zero) global
promotion and adding in the sub-promotion, leading to an $EEI$
of the form\footnote{See appendix~\ref{app:exx}.}
\begin{equation}
\label{eq:exex2}
   EEI = 4 \int \FDRMeasure{q_1} \FDRMeasure{q_2}
        \frac{\hat{\mu}_2^2 (\feynSlash{q}_1 + \feynSlash{k_1})}
             {\FDRbar{q}_1^2 \FDRbar{q}_2^2 \FDRbar{D}_1 \FDRbar{D}_2
          \FDRbar{D}_{12}}=
    i \pi^2 \int \FDRMeasure{q_1}
        \frac{\feynSlash{k_1}}{\FDRbar{q}_1^2 \FDRbar{D}_1}.
\end{equation}
When we move to perform the sub-prescription in the second
sub-divergence, i.e. at fixed $q_2$, we make a similar treatment and find
that the sub-prescription gives an identical contribution due to the
symmetry of the diagram.

\section{Computing {\boldmath $EEI$}s}
\label{app:exx}
Given the important role played by the $EEI$s in the consistency of FDR we explicitly compute the extra-extra integrals in eqs.~(\ref{eq:exex1}) and~(\ref{eq:exex2}):
\bqa
EEI_1 \!&=&\! \int [d^4q_1] [d^4q_2] \frac{\hat \mu^2|_2 (\rlap/q_1 + \rlap/k_1)}{\qbar^4_1\qbar^2_2 \qbar^2_{12} (\qbar^2_1+k_1^2+2 q_1 \cdot k_1)}\nl
EEI_2\! &=&\! \int \FDRMeasure{q_1} \FDRMeasure{q_2}
        \frac{\hat{\mu}_2^2 (\feynSlash{q}_1 + \feynSlash{k_1})}
             {\FDRbar{q}_1^2 \FDRbar{q}_2^2 
          (\qbar^2_1+k_1^2+2  q_1 \cdot k_1 )
          (\qbar^2_2+k_1^2+2  q_2 \cdot k_1 )
          (\qbar^2_{12}+k_1^2+2 q_{12} \cdot k_1)}.\nl
\eqa   
As a first step, we need the related one-loop extra (sub-)integrals
\bqa
I_1(p_1^2)  =  \int [d^4q] \frac{\mu^2}{\qbar^2 \bar D^2_1},~~~
I_2         = \int [d^4q] \frac{\mu^2}{\qbar^2\bar D_1\bar D_2},
\eqa
where $D_i= (q+p_i)^2$ and $\bar D_i= D_i-\mu^2$.
To calculate them we start from the FDR defining expansions of their integrands with $\mu^2 \to q^2$
\bqa
\label{eq:diffex}
\frac{q^2}{\qbar^2 \bar D^2_1} &=&
\left[ 
\frac{q^2}{\qbar^4} 
\right]
-p_1^2
\left[ 
\frac{q^2}{\qbar^6} 
\right]
-2
\left[ 
\frac{q^2 (q\cdot p_1)}{\qbar^6} 
\right]
+4
\left[
\frac{q^2(q \cdot p_1)^2}{\qbar^8} 
\right]+F_1(q),\nl
\frac{q^2}{\qbar^2 \bar D_1 \bar D_2} &=&  \left[ \frac{q^2}{\qbar^6}\right]+F_2(q),
\eqa
where $F_1(q)$ and $F_2(q)$ are UV convergent.
$I_1(p_1^2)$ and  $I_{2}$ are defined~\cite{Pittau:2012zd} as the difference between the l.h.s. of eq.~(\ref{eq:diffex}) and the UV divergent part computed by changing back $q^2 \to \mu^2$ in the numerator
\bqa
I_1(p_1^2) &=& \lim_{\mu \to 0}\mu^2 \int d^nq \left(
\frac{1}{\qbar^2 \bar D^2_1}
-
\frac{1}{\qbar^4} 
+
\frac{p_1^2}{\qbar^6} 
+2
\frac{(q\cdot p_1)}{\qbar^6} 
-4
\frac{(q \cdot p_1)^2}{\qbar^8} 
\right) 
=  -i \pi^2 \frac{p_1^2}{6},\nl
I_2&=& \lim_{\mu \to 0} \mu^2\int d^nq \left(
\frac{1}{\qbar^2 \bar D_1 \bar D_2}-\frac{1}{\qbar^6}
 \right)= \frac{i \pi^2}{2}.
\eqa
Therefore we obtain
\bqa
EEI_1\! &=&\!
\int [d^4q_1] \frac{\rlap/q_1 + \rlap/k_1}{\qbar^4_1(\qbar^2_1+k_1^2+2 q_1 \cdot k_1)}I_1(\bar q_1^2)
=
-\frac{i \pi^2 \rlap/k_1}{12} \int [d^4q_1]
\frac{1}{\qbar^2_1(\qbar^2_1+k_1^2+2 q_1 \cdot k_1)}, 
\nl
EEI_2\! &=&\! \int \FDRMeasure{q_1} 
        \frac{\feynSlash{q}_1 + \feynSlash{k_1}}
             {\FDRbar{q}_1^2  
          (\qbar^2_1+k_1^2+2  q_1 \cdot k_1 )}I_2
= \frac{i \pi^2 \rlap/k_1}{4}\int [d^4q_1]
\frac{1}{\qbar^2_1(\qbar^2_1+k_1^2+2 q_1 \cdot k_1)}.
\eqa
Notice the replacement  $I_1(q_1^2 ) \to I_1(\bar q_1^2)$ in accordance with the global prescription.

Finally, we point out the difference between hatting and not hatting
$\mu^2|_2$, i.e. the inequivalence between $EEI$s and two-loop extra integrals. For instance
\bqa
EEI_1 &=& -\frac{\pi^4 \rlap/k_1}{12}  \left(\ln \frac{k_1^2}{\mur^2}-2 \right), 
\eqa
while
\bqa
 \int [d^4q_1] [d^4q_2] \frac{\mu^2|_2 (\rlap/q_1 + \rlap/k_1)}{\qbar^4_1\qbar^2_2 \qbar^2_{12} (\qbar^2_1+k_1^2+2 q_1 \cdot k_1)}= 
-\frac{\pi^4 \rlap/k_1}{12}  \left(\ln \frac{k_1^2}{\mur^2}+\frac{5}{3}
+\frac{16}{3} f \right). 
\eqa

\bibliography{paper}{}
\bibliographystyle{JHEP}
\end{document}